\documentclass[a4paper,10pt,twoside]{cpc-hepnp}
\usepackage{CJK,upgreek,fancyhdr}
\usepackage{multicol}
\usepackage{graphicx}
\usepackage{booktabs}
\usepackage{amssymb,bm,mathrsfs,bbm,amscd}
\usepackage[tbtags]{amsmath}
\usepackage{lastpage}

\newcommand{\beq}{\begin{equation}}
\newcommand{\eeq}{\end{equation}}
\newcommand{\bea}{\begin{eqnarray}}
\newcommand{\eea}{\end{eqnarray}}

\def\met{\slash{\!\!\!\!\!\!E}_{\text{T}}}
\def\pt{p_{\text{T}}}
\def\to{\rightarrow}

\def\m1{M_1}
\def\m2{M_2}
\def\m3{M_3}

\def\gev{\,{\rm GeV}}

\def\to{\rightarrow}

\newcommand{\lsim}{\mathrel{\mathop{\kern 0pt \rlap
  {\raise.2ex\hbox{$<$}}}
  \lower.9ex\hbox{\kern-.190em $\sim$}}}
\newcommand{\gsim}{\mathrel{\mathop{\kern 0pt \rlap
  {\raise.2ex\hbox{$>$}}}
  \lower.9ex\hbox{\kern-.190em $\sim$}}}

\def\fbi{\,{\rm fb}^{-1}}
\def\abi{\,{\rm ab}^{-1}}

\newcommand{\ee}{{e^{+} e^{-}}}

\newcommand{\scidgts}[2]{\ensuremath{#1\mkern-4mu\times\mkern-4mu 10^{#2}}}

\begin{document}
\begin{CJK*}{GB}{gbsn}

\fancyhead[c]{\small Chinese Physics C~~~Vol. 41, No. 6 (2017) 063102}
\fancyfoot[C]{\small 063102-\thepage}

\footnotetext[0]{Received xx January 2017}

\title{Exotic decays of the 125 GeV Higgs boson at future $e^+ e^-$ colliders
\thanks{ZL is supported by Fermi Research Alliance, 
LLC under Contract No. DE-AC02-07CH11359 with the U.S. Department of Energy. 
LTW is supported by DOE grant DE-SC0013642 and in part by National Science Foundation of China, grant No. 11528509. HZ is supported by IHEP under Contract No. Y6515580U1. }}

\author{%
      Zhen Liu$^{1}$\email{zliu2@fnal.gov}%
\quad Lian-Tao Wang$^{2}$\email{liantaow@uchicago.edu}%
\quad Hao Zhang$^{3}$\email{zhanghao@ihep.ac.cn}%
}
\maketitle

\address{%
$^1$ Theoretical Physics Department, Fermilab, Batavia, IL 60510, USA\\
$^2$ Kavli Institute for Cosmological Physics and the Enrico Fermi Institute, 
The University of Chicago, Chicago, IL 60637, USA\\
$^3$ Institute of High Energy Physics, Chinese Academy of Sciences, Beijing 100049, China\\
}

\begin{abstract}
The discovery of unexpected properties of the Higgs boson would offer an intriguing 
opportunity to shed light on some of the most profound puzzles in particle 
physics. Beyond Standard Model decays of the Higgs boson could 
reveal new physics in a direct manner. Future electron-positron lepton colliders 
operating as Higgs factories, including CEPC, FCC-$ee$ and ILC, with the advantages 
of a clean collider environment and large statistics, could greatly enhance  
sensitivity in searching for these BSM decays. In this work, we perform a general 
study of Higgs exotic decays at future $\ee$ lepton colliders, focusing on the Higgs 
decays with hadronic final states and/or missing energy, which are very challenging 
for the High-Luminosity program of the Large Hadron Collider (HL-LHC). We show 
that with simple selection cuts, $\mathcal{O}(10^{-3}\sim10^{-5})$ limits on the Higgs 
exotic decay branching fractions can be achieved using the leptonic decaying spectator 
$Z$ boson in the associated production mode $\ee\to Z H$. We further discuss the 
interplay between detector performance and Higgs exotic decays, and other 
possibilities of exotic decays. Our work is a first step in a comprehensive study of 
Higgs exotic decays at future lepton colliders, which is a key area of Higgs 
physics that deserves further investigation.  
\end{abstract}



\footnotetext[0]{\hspace*{-3mm}\raisebox{0.3ex}{$\scriptstyle\copyright$}2013
Chinese Physical Society and the Institute of High Energy Physics
of the Chinese Academy of Sciences and the Institute
of Modern Physics of the Chinese Academy of Sciences and IOP Publishing Ltd}%

\begin{multicols}{2}

\section{Introduction}
\label{sec-intro}

The recent discovery of the Higgs boson by the ATLAS and CMS experiments 
at the Large Hadron Collider (LHC) opened a new era of particle physics
\cite{Aad:2012tfa,Chatrchyan:2012xdj}. The Standard Model (SM)-like Higgs 
boson has a very deep connection to many profound puzzles of fundamental physics, 
such as the hierarchy problem and the naturalness problem, the nature of dark matter, 
the origin of neutrino mass, the origin of the fermion mass hierarchy (the flavor problem), 
the essence of the electroweak phase transition and electroweak bayrogenesis. Essentially 
all beyond Standard Model (BSM) solutions to these puzzles, such as supersymmetry (SUSY)~\cite{Flores:1982pr,
Gunion:1984yn,Djouadi:2005gj}, composite Higgs models~\cite{Gripaios:2009pe,
ArkaniHamed:1998nn,Randall:1999ee,Randall:1999vf}, and grand unified theories 
\cite{Georgi:1974sy}, predict modifications of the  properties of the Higgs boson. Hence, 
the precise  measurements of the properties of this Higgs boson have great potential to shed light on BSM physics.

The HL-LHC will measure many SM model decay modes of the Higgs boson to a relative 
precision of $\mathcal{O}(10\%)$ ~\cite{Dawson:2013bba,CMS-NOTE-2013-002,
CMS-DP-2016-064,ATL-PHYS-PUB-2013-014,ATL-PHYS-PUB-2014-016}. The future 
lepton colliders operating as Higgs factories, with their clean collider environment and 
large statistics,  would measure the Higgs boson couplings to a relative precision of 
$\mathcal{O}(0.1\%\sim 1\%)$~\cite{CEPCpreCDR,Fujii:2015jha,Gomez-Ceballos:2013zzn}. 
Many discussions about the physics potential of future lepton colliders focus on the precision 
measurement of the Higgs properties in the effective field theory (EFT) framework 
\cite{Weinberg:1978kz,Buchmuller:1985jz,Grzadkowski:2010es,Ellis:2014jta,Biekoetter:2014jwa,
Contino:2016jqw}. However, new physics could manifest itself through Higgs exotic decays 
if some new light degrees of freedom are present, which are not  described by the SM EFT. 
Hence, systematically searching for Higgs exotic decays  would be an important physics 
component of the future lepton collider programs. Moreover, since many of these future 
facilities are currently at different stages of planning, investigation of these new physics 
potentials could impact their designs to achieve some more comprehensive physics goals.

In Section~2, we present an overview of exotic decay searches at lepton 
colliders and some general discussions of the Higgs exotic decays at different future 
lepton colliders. In Section~3, we describe our simulation framework and 
present our phenomenological analysis for various Higgs exotic decay modes. We 
summarize the physics potential from the Higgs exotic decays at the (HL-)LHC and 
the future lepton collider programs in Section~4. In our summary table, 
we include comprehensive projections and show the complementarity between future 
lepton collider programs and the HL-LHC. We also discuss many important future 
directions for the Higgs exotic decay programs. 

\section{Theoretical framework}
\subsection{Higgs exotic decay modes considered in this work}
The Higgs boson BSM decays have a rich variety of possibilities. To organize this 
study on Higgs boson BSM decays, we selectively choose a set of phenomenologically 
driven processes. We focus on two-body Higgs decays into BSM particles, which are 
allowed to subsequently decay further, up to four-body final states. We only consider 
the Higgs boson as an CP-even particle. CP-violation effects would affect various 
differential distributions, and this demands future study. These processes are well-motivated 
by SM+singlet extensions, two-Higgs-doublet-models, SUSY models, Higgs portals, 
gauge extensions of the SM, etc. These assumptions have also been emphasized in 
the recent overview of Higgs exotic decays~\cite{Curtin:2013fra} and the CERN yellow 
report~\cite{deFlorian:2016spz}.

We consider in general the exotic Higgs decays into BSM particles dubbed as $X_i$, 
$h\to X_1 X_2$. The cascade decay modes are classified into four cases, schematically 
shown in Fig. \ref{fig:topo}. We discuss their major physics motivation and features at 
lepton colliders in order.
\end{multicols}
\begin{center}
\includegraphics[width=16cm]{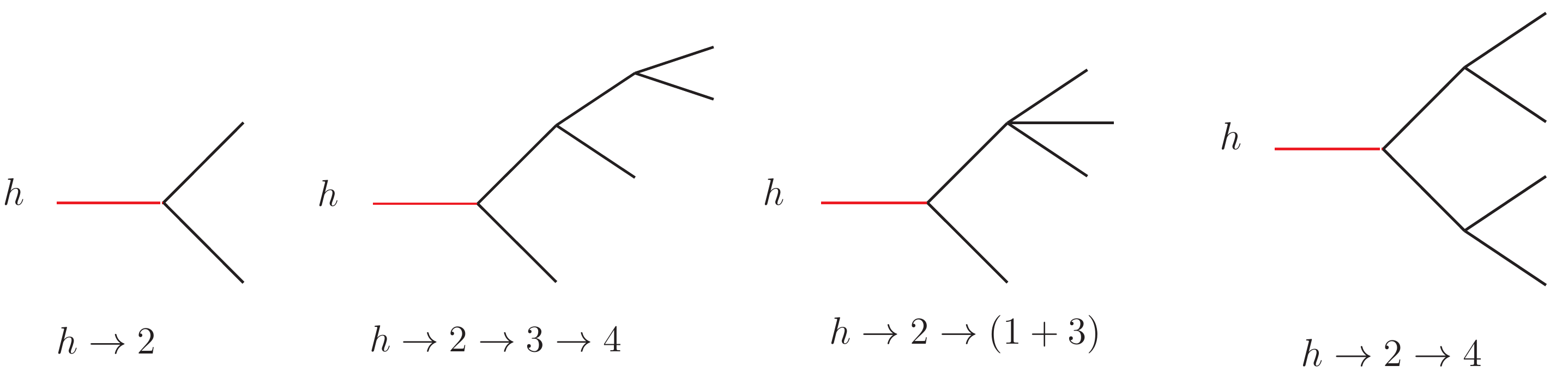}
\figcaption{\label{fig:topo}The topologies of the SM-like Higgs exotic decays. }  
\end{center}
\begin{multicols}{2}
\underline{$h\to2$}: The $X_i$s in this case are detector-stable and charge-neutral.
\footnote{The possibility of a detector-stable electrical charged particle $X_i$ is usually 
more contrived and excluded from direct Drell-Yan production by both LEP and the 
LHC. Hence, we ignore this possibility here.} They could be dark matter candidates. 
The Higgs portal~\cite{Patt:2006fw} to dark matter models, including various SUSY light dark 
matter models~\cite{Arbey:2012na,BhupalDev:2012ru,Han:2013gba,Banerjee:2013fga,
Buckley:2013sca,Hagiwara:2013qya,Belanger:2013pna,Pierce:2013rda,Cao:2013mqa,
Han:2014nba,Huang:2014cla}, motivates this BSM search channel. The lepton collider 
background for this channel are mainly from the process $\ee\to ZZ\to Z+\nu\bar \nu$ 
and $\ee \to W^+W^- \to \ell^+ \ell^- \nu\bar \nu$. This channel, due to its simplicity and 
importance, has been studied by most of the future lepton collider programs~\cite{
CEPCpreCDR,Fujii:2015jha,Gomez-Ceballos:2013zzn} and we will quote these results 
in our summary table. We include this channel here for completeness. In addition, many 
of the models that motivate this channel also induce other Higgs exotic decays we consider in 
this study.

\underline{$h\to2\to3\to4$}: This is the topology in which $X_1$ is detector-stable
and $X_2$ decays to two particles, with one of these decay products further decaying into 
two particles. A typical BSM model for such decay modes is the Higgs decaying into 
the lightest supersymmetric particle (LSP) plus a heavier neutralino, which subsequently 
decays into the LSP plus a resonant BSM particle. This resonant BSM particle could 
be a singlet-like scalar in the Next-to-Minimal-Supersymmetric-Standard-Model (NMSSM). 
Many SUSY models which motivate Higgs invisible decays also induce this decay channel, 
e.g.~\cite{Huang:2013ima,Huang:2014cla}. It also commonly exists in the so-called 
``stealth SUSY'' models~\cite{Fan:2011yu}.  This singlet-like scalar decays into SM fermion 
pairs, giving rise to the final state of  a pair of resonant SM particles plus missing energy, 
dubbed $h\to (ff)+\met$.\footnote{At lepton colliders we could use the quantity missing 
momentum instead of Missing Transverse Energy (MET) $\met$. The former carries more 
information while the latter is more widely used in the hadron collider analyses. For the 
decay channel considered in our analyses, the reach can be improved only marginally by 
the inclusion of the $z$-direction missing momentum information because of the already 
great limit achieved and additional uncertainties from the beamstrahlung effect~\cite{Xiu:2015tha} 
and the initial state radiation (ISR) effect~\cite{Greco:2016izi}. Consequently, we use 
only the more widely adopted missing transverse energy throughout this study.} In this 
study, we only consider the channels which are very challenging at the LHC, $h\to(jj)+\met$, $h\to 
(b\bar b)+\met$ and $h\to (\tau^+\tau^-)+\met$. For the hadronic channels, the major 
background is from the SM Higgs decay modes $h\to ZZ^*\to jj+\nu\bar \nu$ and $h\to 
ZZ^*\to b\bar b+\nu\bar \nu$. For $h\to (\tau^+\tau^-)+\met$, in addition, the SM Higgs decay
$h\to WW^*\to \tau^+\tau^-+\nu\bar\nu$ also contributes to the background.

\underline{$h\to2\to(1+3)$}: This is the topology when $X_1$ is detector-stable
and $X_2$ decays into three particles. The typical BSM model is very similar to the 
previous decay topology. The difference comes from $X_2$ decaying into three 
particles without an intermediate on-shell resonance. This scenario takes place very 
naturally if the singlet-like scalar is heavy, or the particle $X_2$ decays through an off-shell 
Higgs or $Z$-boson. The final state of this Higgs exotic decay signature would be 
$h\to ff+\met$.\footnote{To explicitly distinguish Higgs exotic decays with or without 
resonances in the final state particles, we put the pair of SM particles 
that form a resonance in parenthesis. We follow this notation in describing Higgs exotic decay final 
states  throughout this study.} The lepton collider background sources are similar to 
the previous topology as well.

\underline{$h\to2\to4$}: In this channel, the Higgs decays to a pair of BSM particles, 
both of which subsequently decay into two final state particles. There are a wide range 
of BSM models which give rise to such a decay pattern, and we selectively discuss several 
benchmark cases. The intermediate particle could be a pair of vectors from the ``dark 
photon'' or ``dark Z-prime '' models~\cite{Essig:2013lka}. These models commonly include 
a new gauge boson that couples to the SM with suppressed strength. A typical example 
is a small kinetic mixing with the hypercharge field strengths, but more general couplings 
are certainly possible. They not only induce $h\to Z' Z'$ decays but also sizable $h\to Z Z'$ 
decays in which  the masses of the intermediate particles are uneven, if a source of mass 
mixing is allowed. An important feature of such models relevant for the phenomenology of 
Higgs exotic decay is that the $Z'$ will have sizable $\mathcal{O}({\rm few\%})$ decay branching 
fractions to SM charged leptons. Consequently, this scenario could be severely constrained 
by the (HL-)LHC searches, unless $Z'$ is  leptophobic. This option requires more contrived 
model building to survives collider direct search limits~\cite{Fox:2011qd}. The intermediate 
particle could also be a pair of scalars from SM+scalar, 2HDM+scalar, NMSSM models, 
etc~\cite{Curtin:2013fra}. In addition, a dark sector with strong dynamics, such as the Twin Higgs 
\cite{Chacko:2005pe} models, could also give rise to the Higgs decays into a pair of spin-0 
composite particles.\footnote{However, in many cases, this composite particle ``glu-ball'' 
would be meta-stable and requires a more sophisticated phenomenological study for displaced 
decays.} The intermediate scalar decays strongly prefer SM heavy fermions and thus $b\bar b$, 
$c\bar c$ and $\tau^+\tau^-$ decays would dominate. These decay modes are very hard to 
probe by (HL-)LHC searches due to the large background in the hadron collider environment. 
There is the interesting possibility for the intermediate scalars to decay into diphoton pairs 
that we consider as well. Hence, we consider many combinations of the $(b\bar b)$, $(c\bar c)$, 
$(jj)$, $(\tau^+\tau^-)$ and $(\gamma\gamma)$ decays of the intermediate particles. The 
backgrounds are again mainly from SM Higgs decays into four particles through SM gauge 
bosons. For final states involving photons, the SM electroweak processes at the lepton colliders 
dominate the background. 

There are several other decay topologies in Ref.~\cite{Curtin:2013fra} that we do not include 
in our current study. We comment on them here. \underline{$h\to2\to3$}: In this case, $X_1$ 
is detector-stable and $X_2$ decays promptly. For instance, in dark photon models the SM 
Higgs decays into $h\to \gamma Z'$ and $Z'$ could subsequently decay into SM particles 
via two-body decay. In certain SUSY scenarios, the Higgs could decay to the LSP and the 
next-to-lightest-supersymmetric-particle, which subsequently decays into a photon plus the LSP. 
\underline{$h\to2\to4\to6$} and \underline{$h\to2\to6$}: the direct decay product from the Higgs 
undergoes a decay chain or decays into a three-body final state. These decay topologies are common 
for (R-parity-violating) SUSY models. These decay modes are well-motivated and should be studied 
in follow-up works.
 
\subsection{Higgsstrahlung process}
For future lepton colliders running at the center of mass energy $240\sim250~\gev$, the most 
important Higgs production mechanism is $Z$-Higgs associated production through an off-shell 
$Z$ boson $\ee\to Z^*\to Zh$. The $Z$ boson with visible decays plays the role of Higgs spectator 
and enables Higgs tagging using the ``recoil mass'' technique. Given the known initial state 
energy\footnote{Corrections from beamstrahlung effect~\cite{Xiu:2015tha} and ISR effect~\cite{
Greco:2016izi} need to be carefully taken into account}, subtracting the $Z$-boson four-momentum 
enables the reconstruction of the Higgs four-momentum and thus its invariant mass. The so called 
recoil mass defined in this way sharply peaks at the Higgs boson mass. A selection cut around 
this peak would remove the majority of the SM background.

For an unpolarized electron-positron beam at the center of mass energy $240~\gev$, the Higgs 
production rate is around 230~fb~\cite{Sun:2016bel,Gong:2016jys}. Both the CEPC and FCC-$ee$ 
plan to mainly run with unpolarized beam at this energy. Here, we consider the CEPC running 
scenario with an effective integrated luminosity of $5~\abi$, following its current plan of two 
interaction points and ten years of running with the designed beam luminosity. The CEPC will 
produce $1.15$ million Higgs bosons in this production channel. The FCC-$ee$ running scenario 
we consider here has six times more statistics than the CEPC, $30~\abi$ (equivalently 6.9 million 
Higgs bosons), following its current plan of four interaction points with a higher beam luminosity. 
For the ILC, we only consider the limits from its $250~\gev$ runs, in the H20 scenario of $2~\abi$ 
integrated luminosity with beam polarization of $p(e^-,e^+)=(-0.8,+0.3)$. The Higgs production 
rate is enhanced by a factor $1.4$ due to beam polarization compared to the unpolarized beams 
of circular lepton colliders. In this scenario, considering the 250~\gev\ runs alone, the ILC will 
produce $0.64$ million Higgs bosons in this channel.

Before proceeding to the next section of detailed numerical analysis for individual channels, we 
comment on several instructive scenarios for the future lepton collider sensitivities here. If we 
consider the cleanest $\ee\to Zh, Z\to \ell^+\ell^-$ mode, the Higgs boson can be tagged with 
little background from the SM. Taking this leptonic-decaying spectator $Z$-boson alone and 
CEPC as an example, we will have $7.7\times 10^{4}$ Higgs bosons, naively reaching a very 
impressive $4\times 10^{-5}$~($2.5\times 10^{-4}$) limit on the Higgs exotic branching fraction 
for the case of zero (one hundred) SM background. In this study, we choose to study this clean 
leptonic spectator $Z$-boson mode with various Higgs exotic decay modes. For most Higgs 
exotic decay modes, further inclusion of hadronic decaying $Z$ boson and even invisible $Z$ 
will definitely improve the limits significantly. The (HL-)LHC will produce more Higgs bosons, 
providing excellent limits on Higgs decaying into leptons, such as $h\to(\ell^+\ell^-)(\ell^+\ell^-)$, 
reaching better than $\mathcal{O}(10^{-5})$ branching fractions. Hence, we do not consider these pure 
leptonic channels at future lepton colliders but focusing only on the channels that are hard for the 
LHC, involving hadronic decays and/or missing energy.

\section{Phenomenological analysis}
Following the discussion in the previous section, we perform a numerical study of the future 
lepton collider reach for selected Higgs exotic decay modes. Though operating at slightly different 
center of mass energies in the range of $240\sim250~\gev$, the proposed future lepton colliders in general have 
similar detector performance. For simplicity, we choose the CEPC as the benchmark accelerator 
and detector model in this section for the analysis. We will extrapolate the sensitivity for FCC and 
ILC in the next section.

For numerical analyses, we generate both the signal and the background events for an 240 GeV 
electron-positron collider with MadGraph5 at parton level \cite{Alwall:2014hca} and impose the 
detector acceptance, energy and momentum smearing, and separation cuts with our own analysis 
code.

We describe here our parameter choices for the detector effects, and our pre-selection cuts that 
are universal for the analyses for all Higgs exotic decay mode. To reach a high particle identification 
efficiency, all of the visible particles in the final state are required to have $|\cos\theta|<0.98$ (following 
Ref.~\cite{CEPCpreCDR}), or equivalently $|\eta|<2.3$. The final state particles are required to be 
well separated with 
\beq
y_{ij}\equiv\frac{2\min\left(E_i^2,E_j^2\right)\left(1-\cos\theta_{ij}\right)}{E_{vis}^2}\geqslant0.001. 
\label{eq:separationcut}
\eeq
We only study the case where the $Z$ boson decays into $\ell^+\ell^-$
final state where $\ell^\pm=e^\pm, \mu^\pm$, and leave the study of other decay modes of the 
$Z$ boson for future works. The signal events are required to contain at least a pair of 
opposite-sign same-flavor charged leptons with an opening angle greater than 
$80^\circ$, and satisfy
\beq
E_{\ell}>~5~{\text{GeV}}
\eeq
and
\beq
|m_{\ell\ell}-m_Z|<10~{\text{GeV}}.
\eeq
Furthermore, the recoil mass is defined as
\beq
m_{\text{recoil}}\equiv\sqrt{s-2\sqrt{s}E_{\ell\ell}+m_{\ell\ell}^2}
\eeq
where $E_{\ell \ell } = E_{\ell^+} + E_{\ell^-}$.
The recoil mass is required to satisfy
\beq
\left|m_{\text{recoil}}-m_h\right|< 5~{\text{GeV}}.
\label{eq:recoilcut}
\eeq

To suppress the ISR contribution to the backgrounds, for Higgs exotic decay modes
without missing energy, we require the events to have the total visible energy
\beq
E_{vis}>~225~{\text{GeV}}.
\eeq

In this work, we mimic the detector resolution effect by adding Gaussian smearing effects 
on the four-momentum of the particles, following the performance described in Ref.~\cite{CEPCpreCDR}. 
For photons in the final state, the energy resolution is determined by the electromagnetic 
calorimeter, which performs approximately as
\beq
\frac{\delta E}{E}=\frac{0.16}{\sqrt{E/{\text{GeV}}}}\oplus0.01.
\eeq
The energy resolution of jets is affected by the hadron calorimeter, and performs approximately as
\beq
\frac{\delta E}{E}=\frac{0.3}{\sqrt{E/{\text{GeV}}}}\oplus0.02.
\eeq
For electrons and muons in the final state, we include the momentum resolution effect of the 
track system with the approximate performance of
\beq
\Delta\left(\frac{1}{\pt}\right)=2\times10^{-5}\oplus \frac{10^{-3}}{\pt\sin\theta}.
\eeq

Next, we discuss the phenomenology of individual Higgs exotic decay channels.

\subsection{$h\to (jj)+\met $}
This final state appears if the SM-like Higgs boson decays into $X_2X_1$
with $X_2\to X_1 s$ and $s\to j j$. In the NMSSM, for example, the particles $X_1$, 
$X_2$ and $s$ could be identified as the light neutralinos $\tilde \chi_1^0$, $\tilde \chi_2^0$ and 
the light singlet-like (pseudo-)scalar $h_1 (a_1)$, respectively. We generate 
the irreducible SM background $e^+e^-\to \ell^+\ell^-\nu_\ell\bar\nu_\ell j j$
with MadGraph5. Beyond the pre-selection cut and the recoil mass cut, 
we require that there are two additional jets which satisfy
\beq
E_j>~10~{\text{GeV}}~{\rm and}~|\cos\theta_j|<0.98.
\eeq
After these cuts, the invariant mass distribution of the 
dilepton system is shown in Fig.~\ref{fig:zvvjj}. 
\begin{center}
\includegraphics[width=8cm]{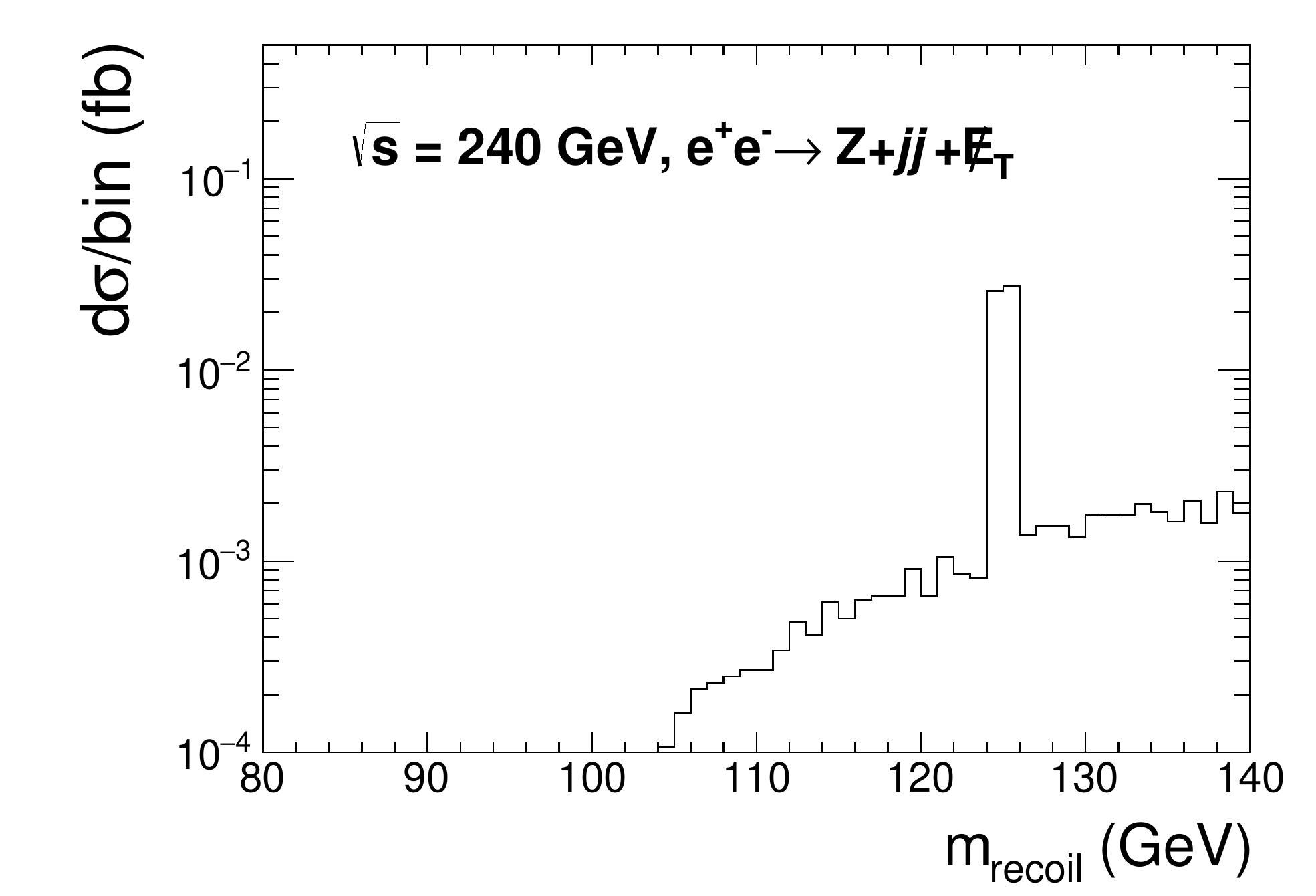}
\figcaption{\label{fig:zvvjj}The recoil mass distribution of the SM backgrounds for 
$\ell^+\ell^-\nu_\ell\bar\nu_\ell j j$.  
All of the preliminary
cuts except the recoil mass cut are applied.}  
\end{center}
The dominant 
background after the recoil mass cut will clearly be the Higgsstrahlung process with 
$h\to ZZ^* \to q\bar q\nu\bar\nu$. 
After the recoil mass cut, 
the SM background cross section is 0.063~fb.
\begin{center}
\includegraphics[width=8cm]{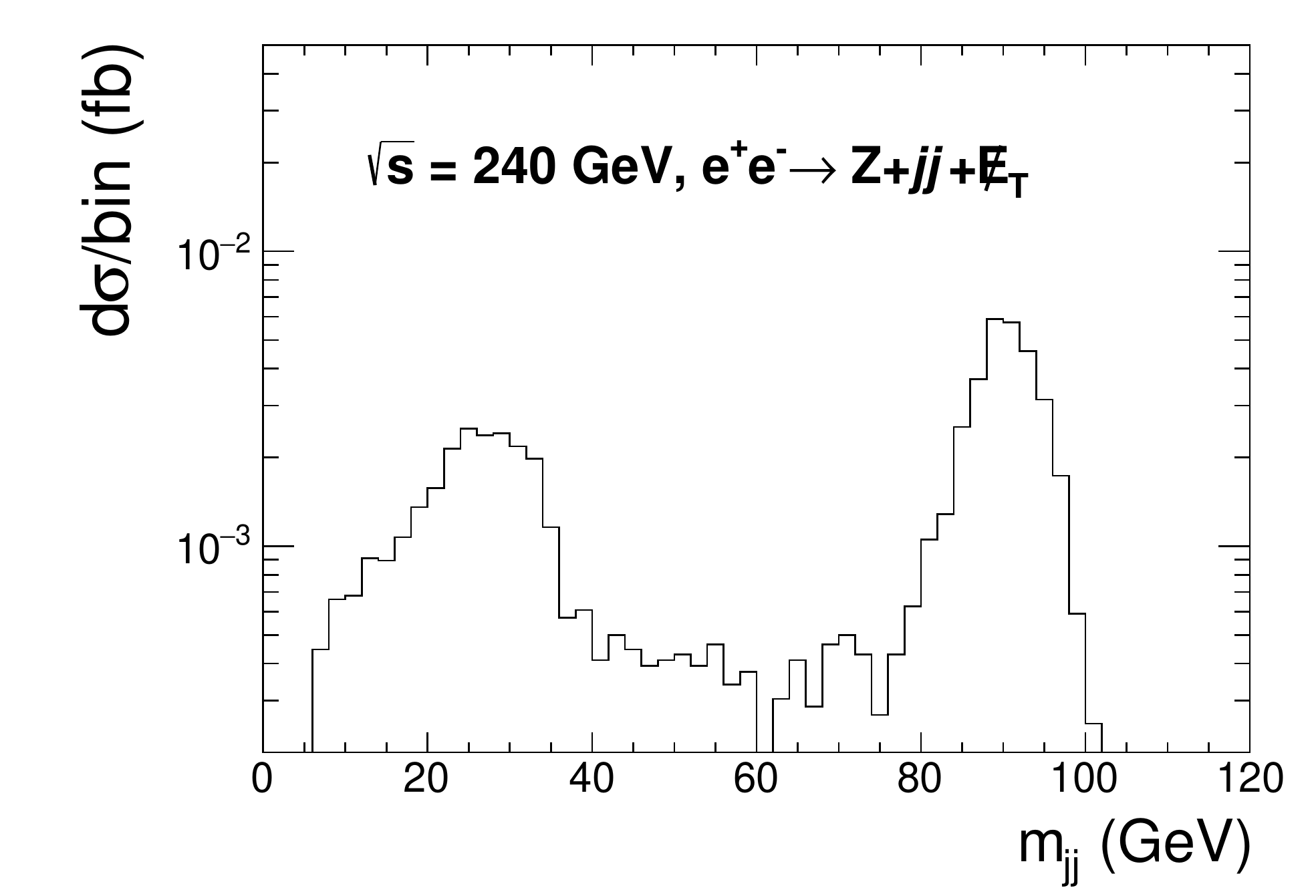}
\includegraphics[width=8cm]{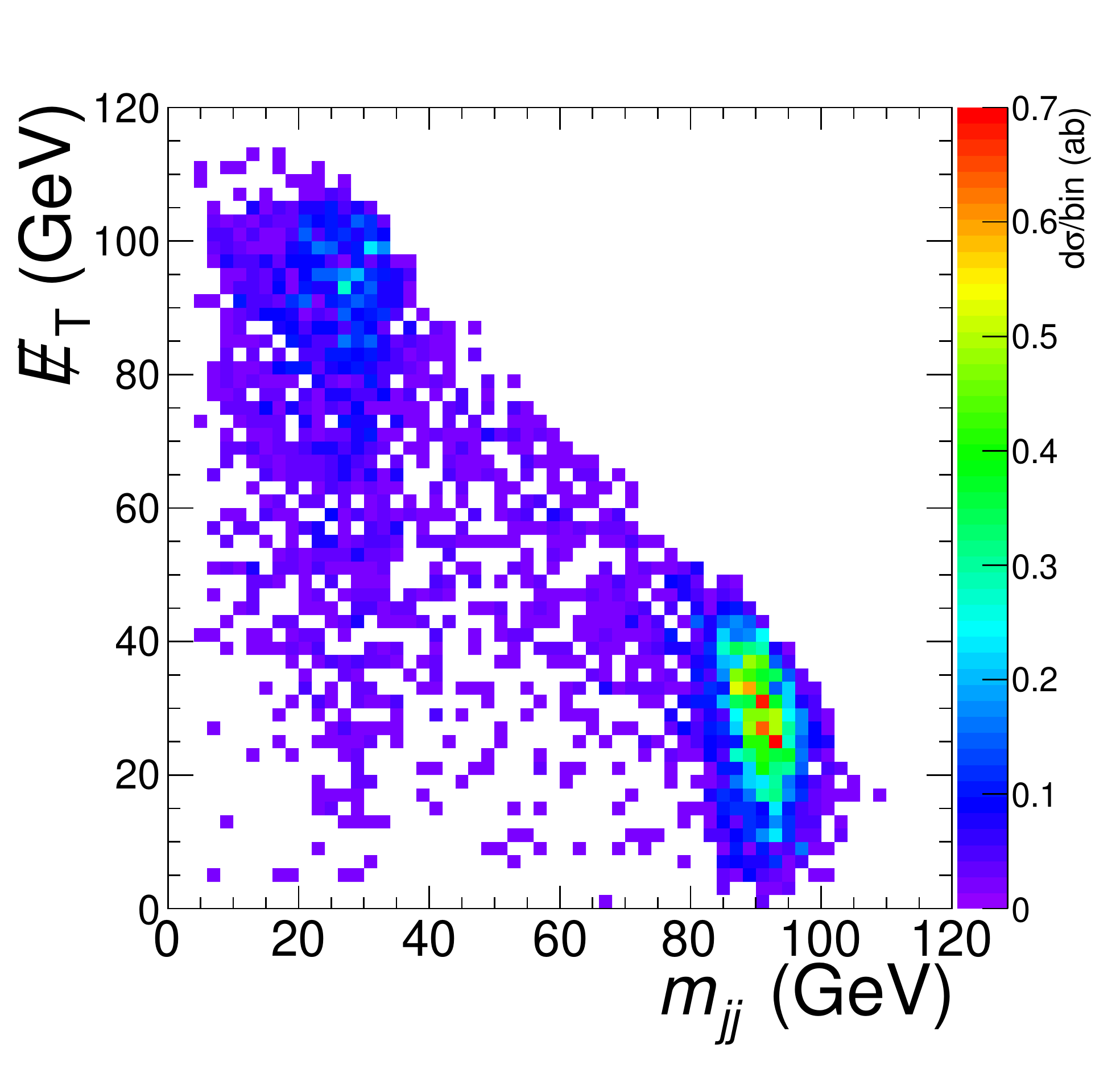}
\figcaption{\label{fig:zvvjj2}The invariant mass distribution of the SM backgrounds for 
$\ell^+\ell^-\nu_\ell\bar\nu_\ell j j$ (a), and the 
$m_{jj}$-$\met$ distribution of the SM backgrounds for 
$\ell^+\ell^-\nu_\ell\bar\nu_\ell j j$ (b) after cuts. }  
\end{center} 

The dijet invariant mass ($m_{jj}$) distribution and 
the two-dimensional differential distribution of  $m_{jj}$ versus $\met$ 
of the SM background after the recoil mass
cut are shown in Fig. \ref{fig:zvvjj2}. There is a clear valley in the distribution between
35 to 75 GeV, in which none of the $Z$ bosons from the SM-like Higgs boson decay
are on-shell and thus the $h\to q\bar q\nu_\ell\bar\nu_\ell$ is doubly suppressed. 
This property could be used to optimize the cut and increase the sensitivity 
to the signal if the  invariant mass of the light (pseudo)scalar falls in this range.

We use the likelihood function of the $m_{jj}$-$\met$ distribution to give the 
exclusion limit. We show the 95\% C.L. exclusion limit in Fig.~\ref{fig:zvvjj3} in the 
plane of $X_1$, mass $m_1$, and the mass splitting between $X_2$ and $X_1$,  
$m_2-m_1$, for two benchmark intermediate scalar masses of 10~\gev\ (a) 
and 40~\gev\ (b). In most of the parameter space, the Higgs branching fraction 
to this exotic decay channel $h\to (jj)+\met$ can be excluded to the level of 2-6$\times 
10^{-4}$. We discuss several kinematical features here that impact the exclusion limit.
When $m_{s}=10$~GeV, the larger the $m_{2}-m_{1}$, the better the reach. This is 
due to the fact that the signal events populate the lower-left corner in the $m_{jj}$-$\met$ 
plane with low SM background for large mass splitting when the MET is small. 
Consequently, the highest sensitivity is reached when $m_{1}=10~{\text{GeV}}, 
m_{2}=100~{\text{GeV}}$. When $m_{2}-m_{1}$ is small, the signal events will tend to 
have a large $\met$ and look like the SM background events $e^+e^-\to Z(h\to ZZ^*)$
where the on-shell $Z$ from the Higgs boson decay decays to $\nu\bar\nu$. Furthermore, 
for a light intermediate scalar mass, small mass splitting also results in soft jets in the final 
states which are more likely to fail the pre-selection cuts of two jets.
When $m_{h_1}=40$ GeV (so $m_{2}-m_{1}\geqslant40$ GeV), the $m_{jj}$ falls 
in the valley and the sensitivity is high (around $2\times 10^{-4}$)
as we expected. When the intermediate scalar mass $m_{h_1}$ is close to the $Z$ boson mass, 
the sensitivity is low, due to the relatively 
large SM background $e^+e^-\to Z(h\to ZZ^*)$
where the on-shell $Z$ from the Higgs boson decay decays to dijets.
\end{multicols}
\begin{center}
\includegraphics[width=7.5cm]{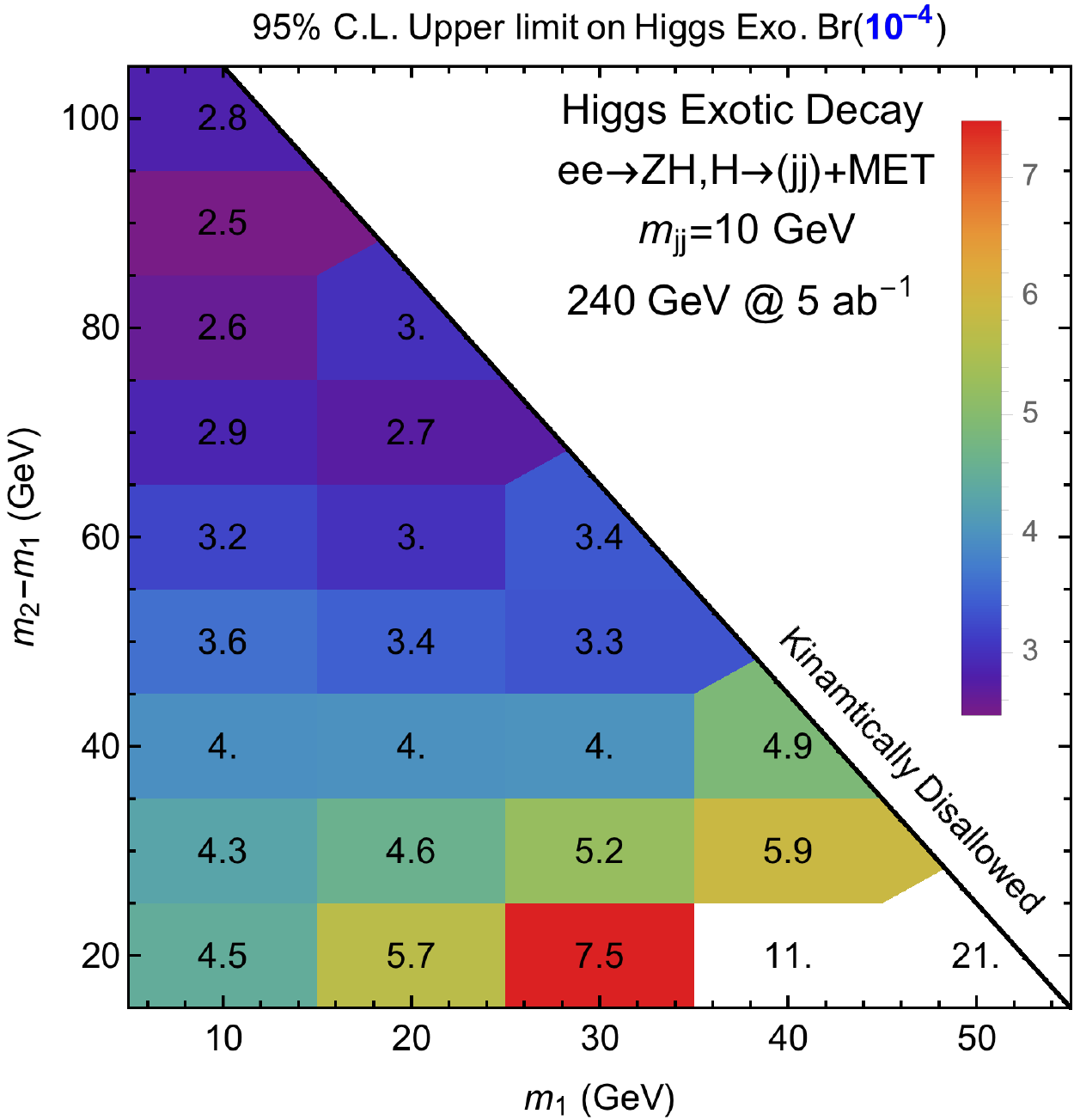}\quad\quad\quad\quad\quad
\includegraphics[width=7.5cm]{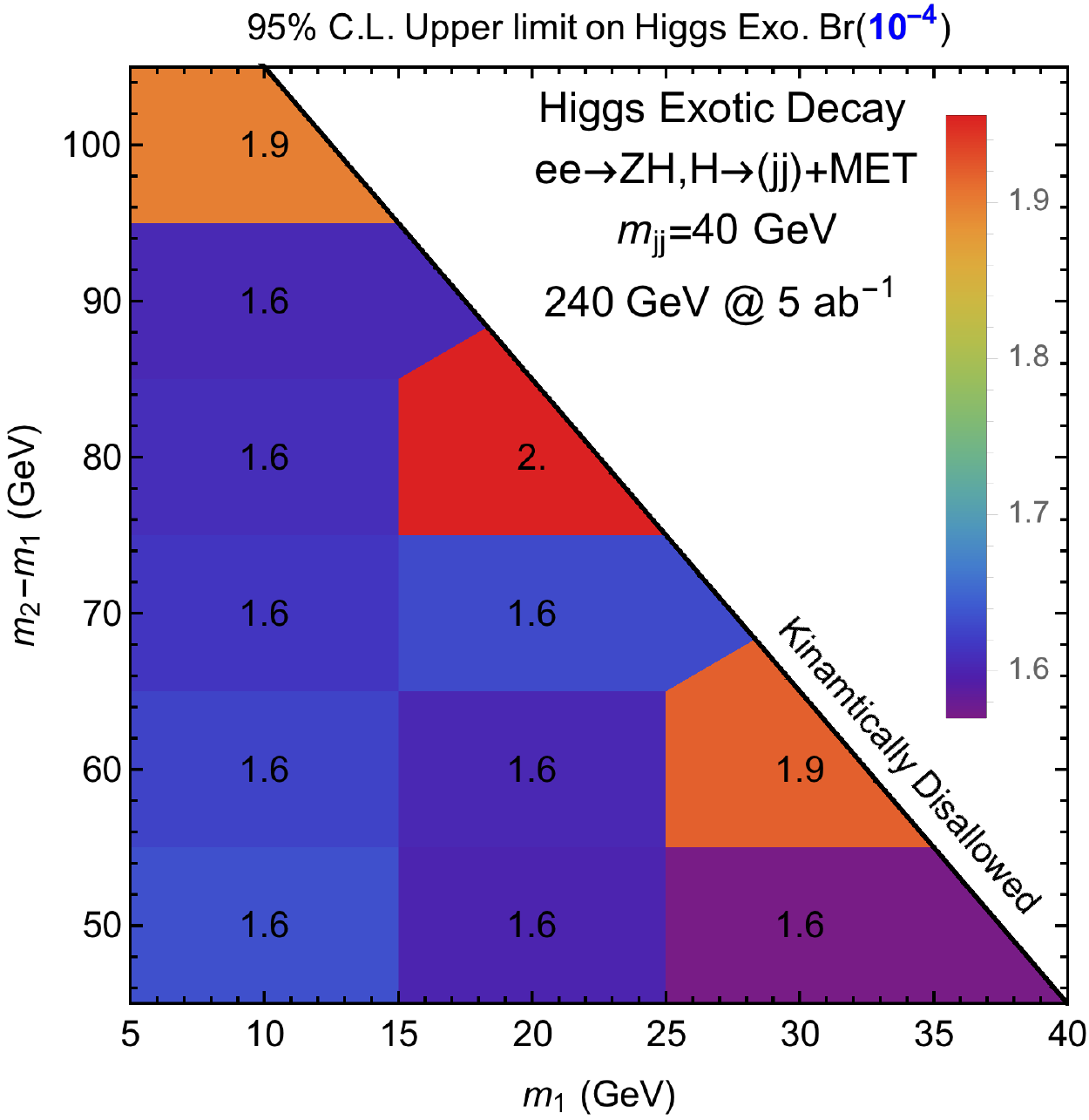}
\figcaption{\label{fig:zvvjj3}The 95\% C.L. upper limit on the Higgs exotic decay branching fractions into 
$(jj)+\met$ for various lightest detector-stable particle $X_1$ with mass $m_1$ and mass splittings 
$m_2-m_1$. The results for the benchmark cases of the dijet mother particle mass of 10~GeV 
and 40~GeV are shown in (a) and (b), respectively.}  
\end{center}
\begin{multicols}{2}

\subsection{$h\to (b\bar b)+\met $}
The background and the benchmark model for this mode are the 
same as the $h\to (jj)+\met $ case discussed in the previous section. The signal event is 
required to contain two $b$-tagged
jets. The $b$-tagging efficiency is conservatively chosen to be 80\%,
and the charm mis-tagging rate and the light flavor mis-tagging rate are set to be 9\%
and 1\%, respectively. Similarly, we use the likelihood function of the 
$m_{b\bar b}$-$\met$ distribution to derive the 
exclusive limit. 

We show the 95\% C.L. exclusion limit in Fig.~\ref{fig:zvvjj4} in the plane of $X_1$, mass $m_1$, 
and the mass splitting between $X_2$ and $X_1$, $m_2-m_1$, for two benchmark intermediate 
scalar masses of 10~\gev\ (a) and 40~\gev\ (b). In most of the parameter space, the 
Higgs branching fraction to this exotic decay channel $h\to (b\bar b)+\met$ can be excluded to 
the level of $5\times 10^{-5}\sim 1.5\times 10^{-4}$. The features for various kinematical regions 
are also similar to the analysis in the previous section. The limits are roughly a factor of 4 better 
than $h\to (jj)+\met$ across the whole parameter region, due to the $b$-tagging reducing the 
flavor universal quark jets background. In the highest sensitivity benchmark points, the 95\% C.L. 
exclusion bound can reach $6\times 10^{-5}$ (e.g., $m_{1}=10~{\text{GeV}}, m_{2}=~60\sim
100~{\text{GeV}}, m_{s}=40~{\text{GeV}}$). This impressive result nearly reaches 
the statistical limit of CEPC, after folding in a factor 0.64 on signal strength from the requirement 
of double $b$-tagging.
\end{multicols}
\begin{center}
\includegraphics[width=7.5cm]{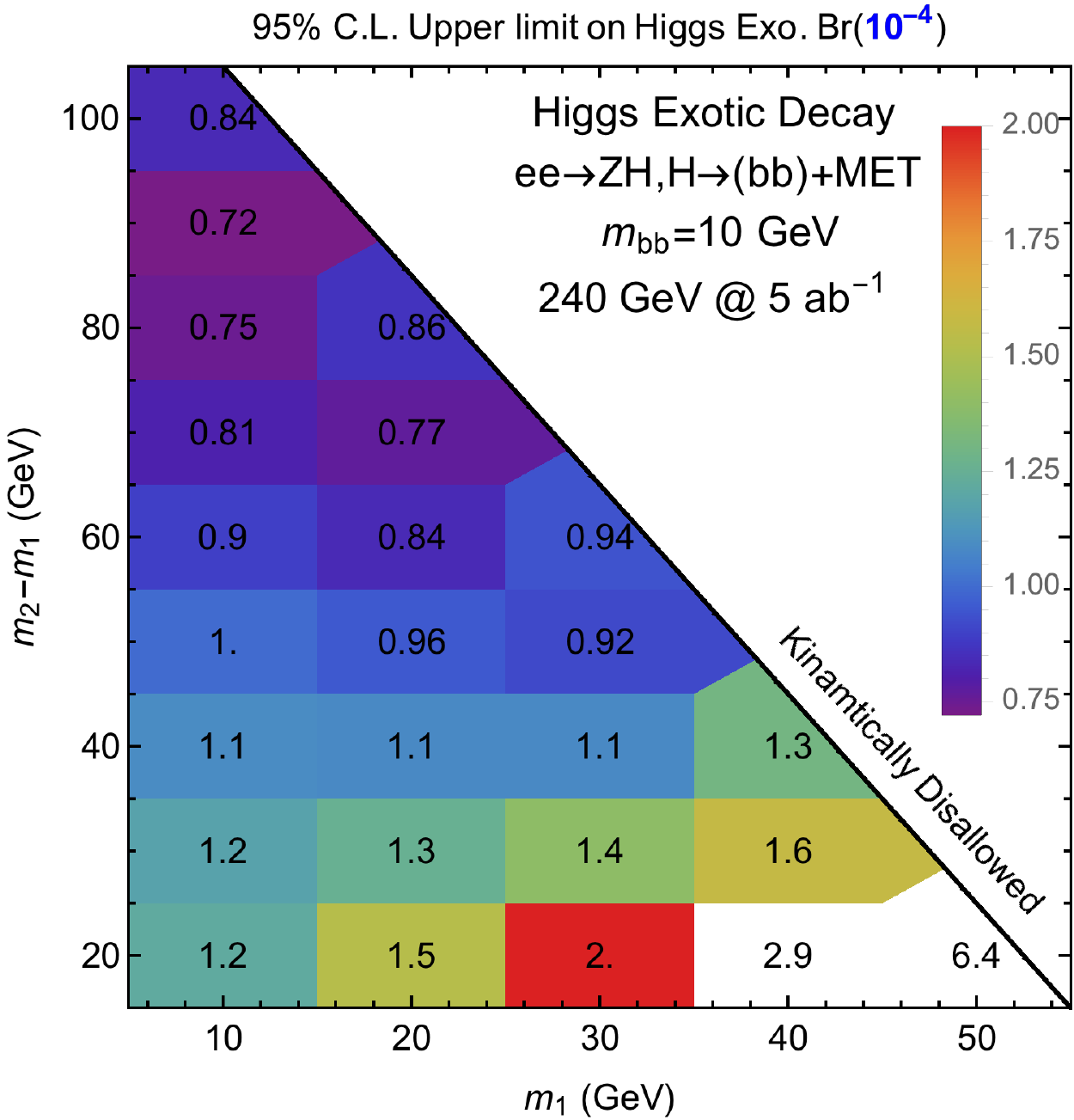}\quad\quad\quad\quad\quad
\includegraphics[width=7.5cm]{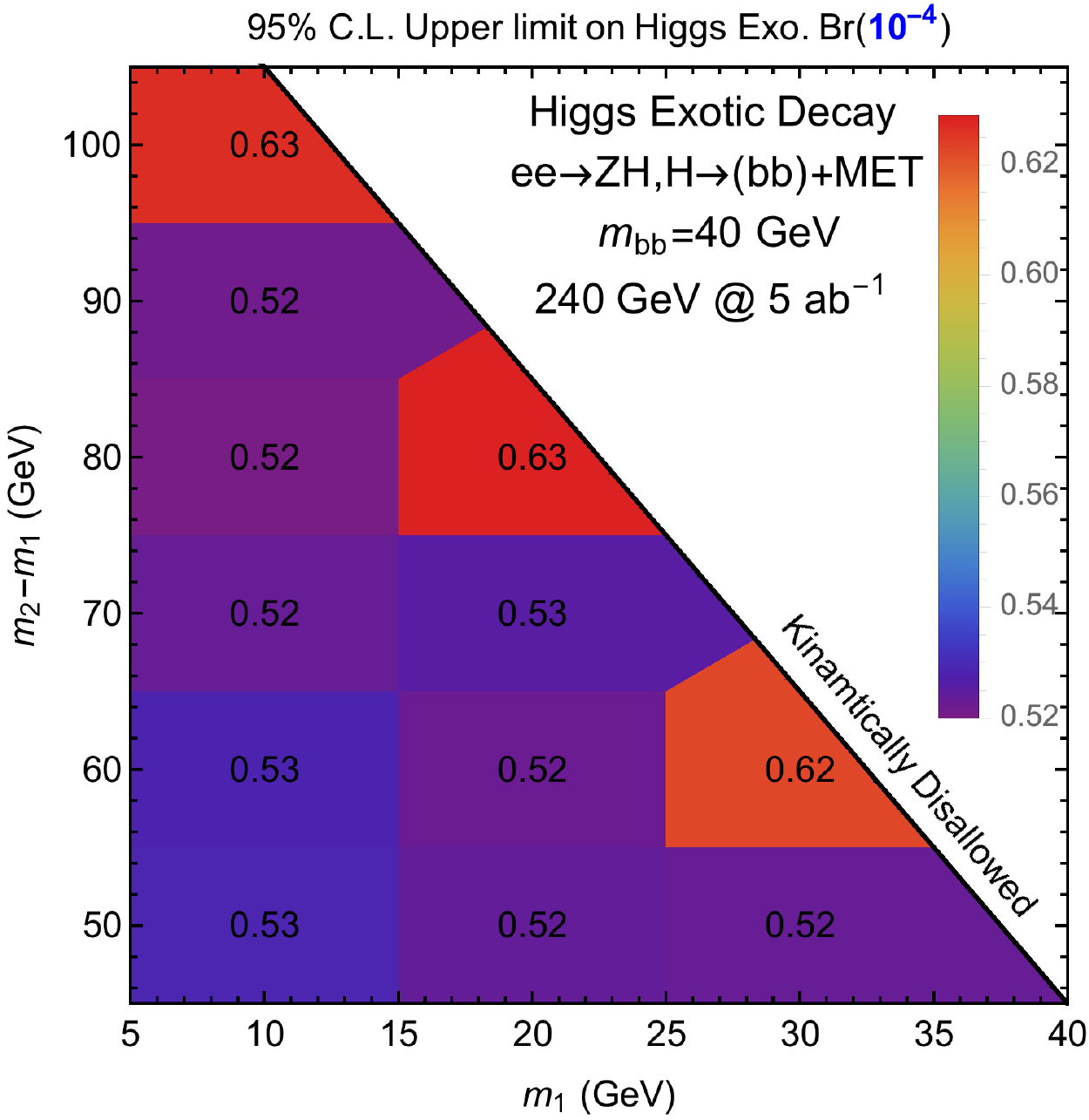}
\figcaption{\label{fig:zvvjj4}The 95\% C.L. upper limit on the Higgs exotic decay branching 
fractions into $(jj)+\met$ for various lightest detector-stable particle with mass $m_1$ and mass 
splittings $m_2-m_1$. The results for the benchmark cases of a dijet mother particle mass 
of 10~GeV and 40~GeV are shown in (a) and (b), respectively.}  
\end{center}
\begin{multicols}{2}

\subsection{$h\to jj+\met $ and $h\to b\bar b+\met$}
Although the final states of these channels are the same as the $h\to(jj)+\met$  and $h\to (b\bar b)+\met$ 
cases discussed in the previous sections, the distributions 
of the kinetic variables are quite different since there is no dijet resonance in this case.
We assume the intermediate particle (e.g., scalar $s$) to be very heavy so that the decay of the 
$X_2$ can be fully described by a four-fermion contact operator. 
Similarly, we use the likelihood function of the 
$m_{b\bar b}$-$\met$ distribution to derive the 
exclusive limit. The results are
shown in Fig.~\ref{fig:zvvjj5} in the plane of $X_1$, mass $m_1$, and the mass splitting 
between $X_2$ and $X_1$, $m_2-m_1$, for $h\to jj+\met$ (a) and $h\to b\bar b 
+\met$ (b). Comparing with the Higgs exotic decays with intermediate resonance 
in the previous section, the exclusion limits on the branching fraction are only slightly worse 
in the bulk region of the parameter space, reaching generally $3\times10^{-4}\sim 8\times 
10^{-4}$ and $2\times10^{-4}\sim4\times 10^{-4}$ for $h\to jj+\met$ and $h\to b\bar b+\met$, 
respectively.
\end{multicols}
\begin{center}
\includegraphics[width=7.5cm]{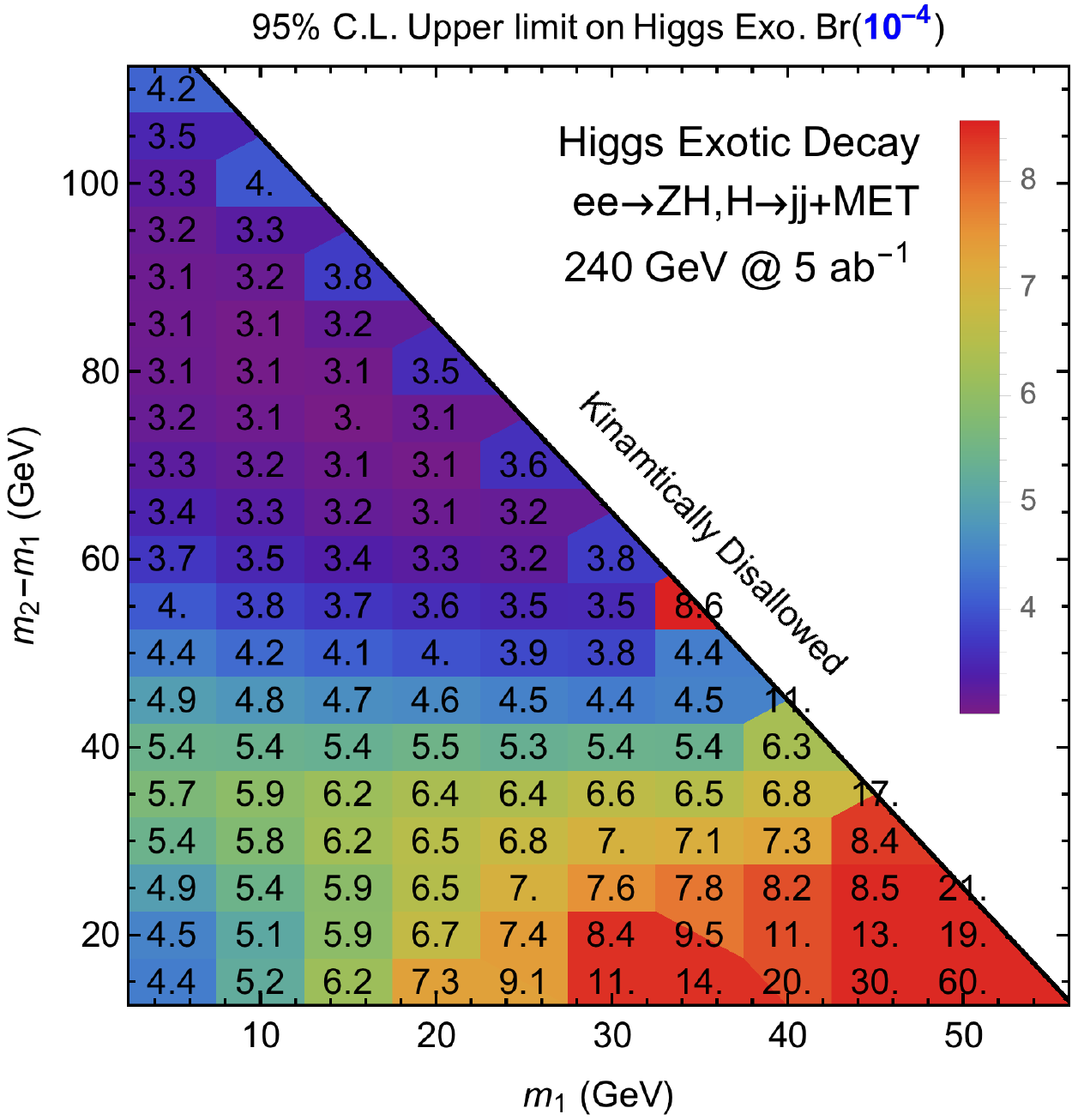}\quad\quad\quad\quad\quad
\includegraphics[width=7.5cm]{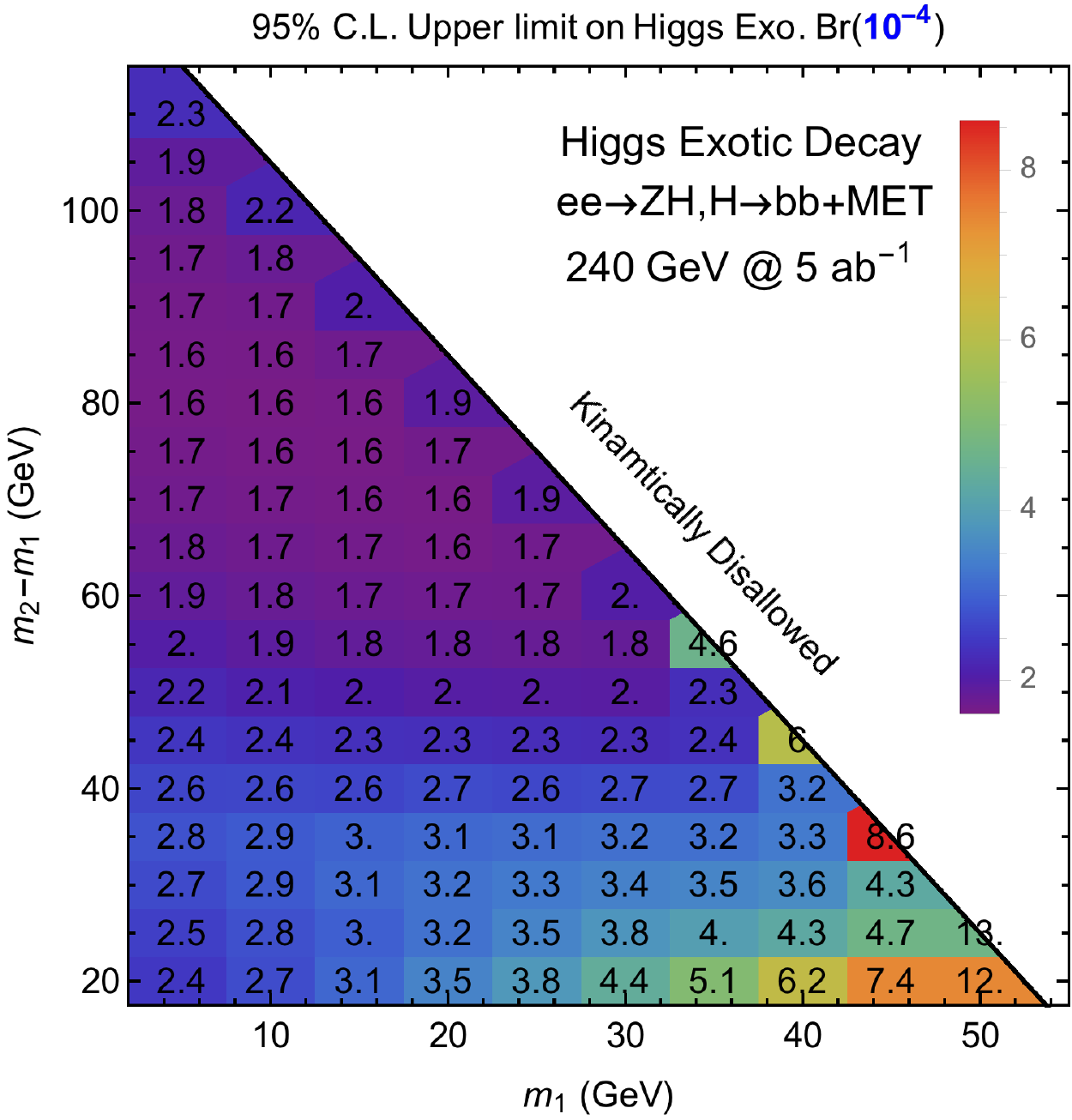}
\figcaption{\label{fig:zvvjj5}The 95\% C.L. upper limit on the Higgs exotic decay branching 
fractions into $jj+\met$ (a) and $b\bar b + \met$ (b) for various lightest 
detector-stable particle mass $m_1$ and mass splittings $m_2-m_1$.}  
\end{center}
\begin{multicols}{2}

From the exclusion limits shown in Fig.~\ref{fig:zvvjj5}, we find that when the mass splitting 
$m_2-m_1$ is around 80~GeV, the future lepton colliders have the strongest sensitivities 
on these Higgs exotic channels,  reaching around $3.1\times 10^{-4}$ and $1.6\times 10^{-4}$ 
for $h\to jj +\met$ and and $h\to b\bar b + \met$, respectively.  
When $X_1$ is light and $m_2-m_1$ is large, the energy is shared by the two jets and the $X_1$.
Consequently, when the mass splitting $m_2-m_1$ is around 80~GeV, the dijet invariant mass 
will be around 40$\sim$60~\gev,  falling in the ``valley'' of low SM background as shown in Fig.~\ref{fig:zvvjj2}. 
For heavier $X_1$, the MET will be lower due to less momentum available for the LSP. The 
optimal limits will be reached for an even smaller mass splitting. The $b\bar b
+\met$ case has a higher sensitivity again by roughly a factor of two since the $b$-tagging suppresses the 
SM background.

\subsection{$h\to(jj)(jj)$, $h\to(c\bar c)(c\bar c)$ and $h\to(b\bar b)(b\bar b)$}
For this class of Higgs exotic decays, we consider the scalar mediator~($s$), the pseudoscalar~($a$) and the vector~($Z^{\prime \mu}$) mediator. We assume the effective interactions 
between the SM-like Higgs boson and the mediators are $hss$, $haa$ and 
$hZ^{\prime \mu} Z_\mu^\prime$, respectively. The (pseudo)scalar mediator can decay into 
dijet final states via $s\bar ff$ ($a\bar f \gamma_5f$) or $sG_{\mu\nu}G^{\mu\nu}$
($aG_{\mu\nu}\tilde G^{\mu\nu}$) interactions.
For the vector mediator case, we consider both the vector-like and right-handed 
interaction with the SM fermions.
\begin{center}
\includegraphics[width=8cm]{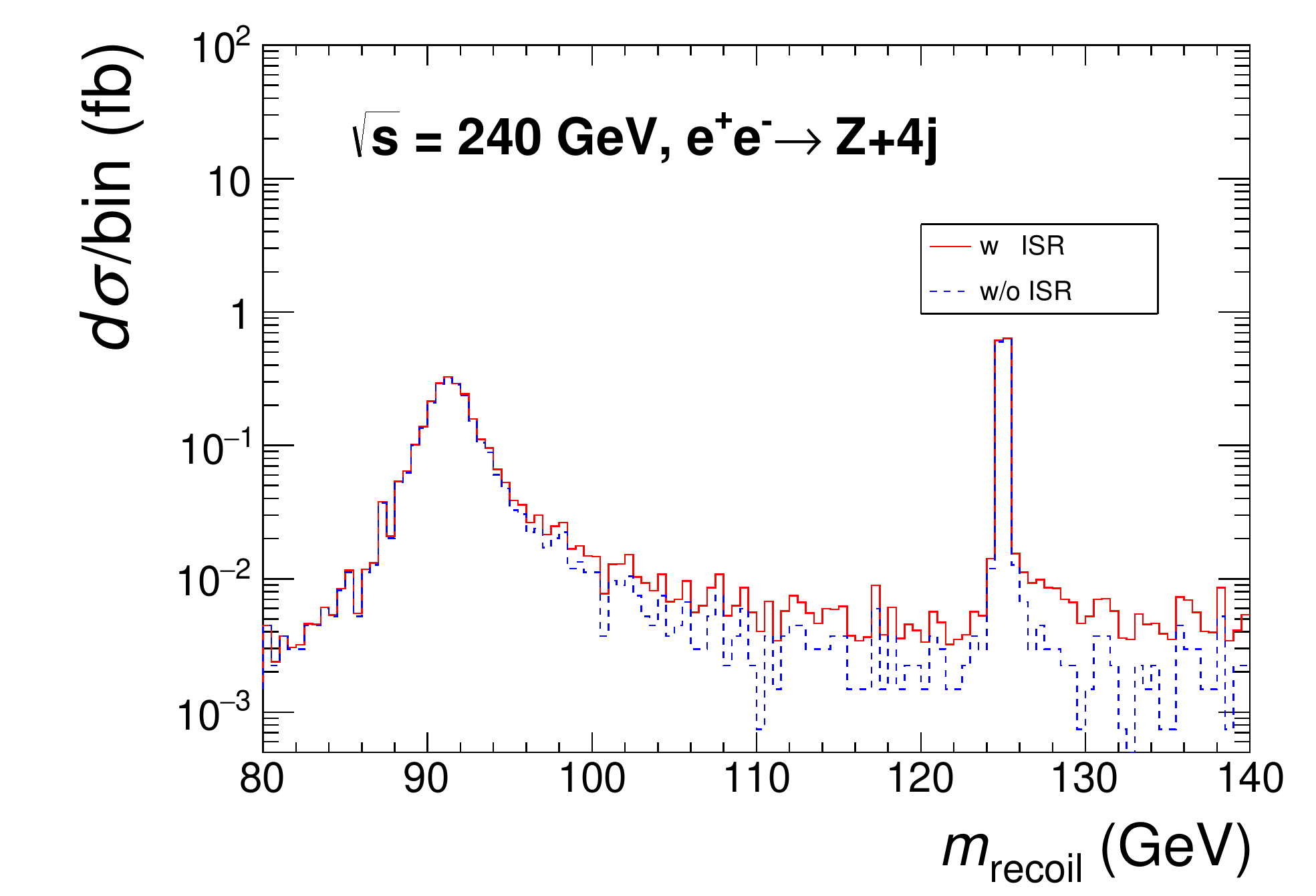}
\figcaption{\label{fig:zjjjj}The recoil mass distribution of the SM backgrounds for $Z+4j$.
All cuts except for the recoil mass cut are applied. The red curve contains the contribution from 
$e^+e^-\to Z+4j+\gamma$.}  
\end{center}

In addition to the pre-selection cuts and the recoil mass cut, we require that
there are at least four jets that satisfy 
\beq
E_j>5~\gev.
\eeq

The most important background from the SM is 
\beq
e^+e^-\to Zh\nonumber
\eeq
with
\beq
Z\to \ell^+\ell^-, h\to j j j j,
\eeq
where the four jets could be either from the hadronic decay of the SM vector bosons in 
$h\to VV^*$, or from $h\to jj$ with jet-splitting. We show the recoil mass distribution of 
the SM background with all but the recoil mass cut applied in Fig.~\ref{fig:zjjjj}. We also 
include in the red curves the background distribution with the inclusion of ISR effect. Its effect is negligibly small after the relatively large window of the recoil 
mass cut. We hence neglect the ISR effect in this analysis. 

Another kinematical variable which is useful for separating the signal and background is
\beq
\delta m\equiv\min_{\sigma\in A_4}\left|m_{j_{\sigma(1)}j_{\sigma(2)}}
-m_{j_{\sigma(3)}j_{\sigma(4)}}\right|.
\eeq
With the combination which gives $\delta m$, we calculate the likelihood function of the 
$m_{j_1j_2}+m_{j_3j_4}$ versus $\delta m$ distribution and get the 95\% C.L. exclusive bound
shown in Fig. \ref{fig:brjjjj1}. The chiral structure does not affect the result significantly. 
In the exclusion bounds, we only show the vector current result of the vector mediator. 

\begin{center}
\includegraphics[width=8cm]{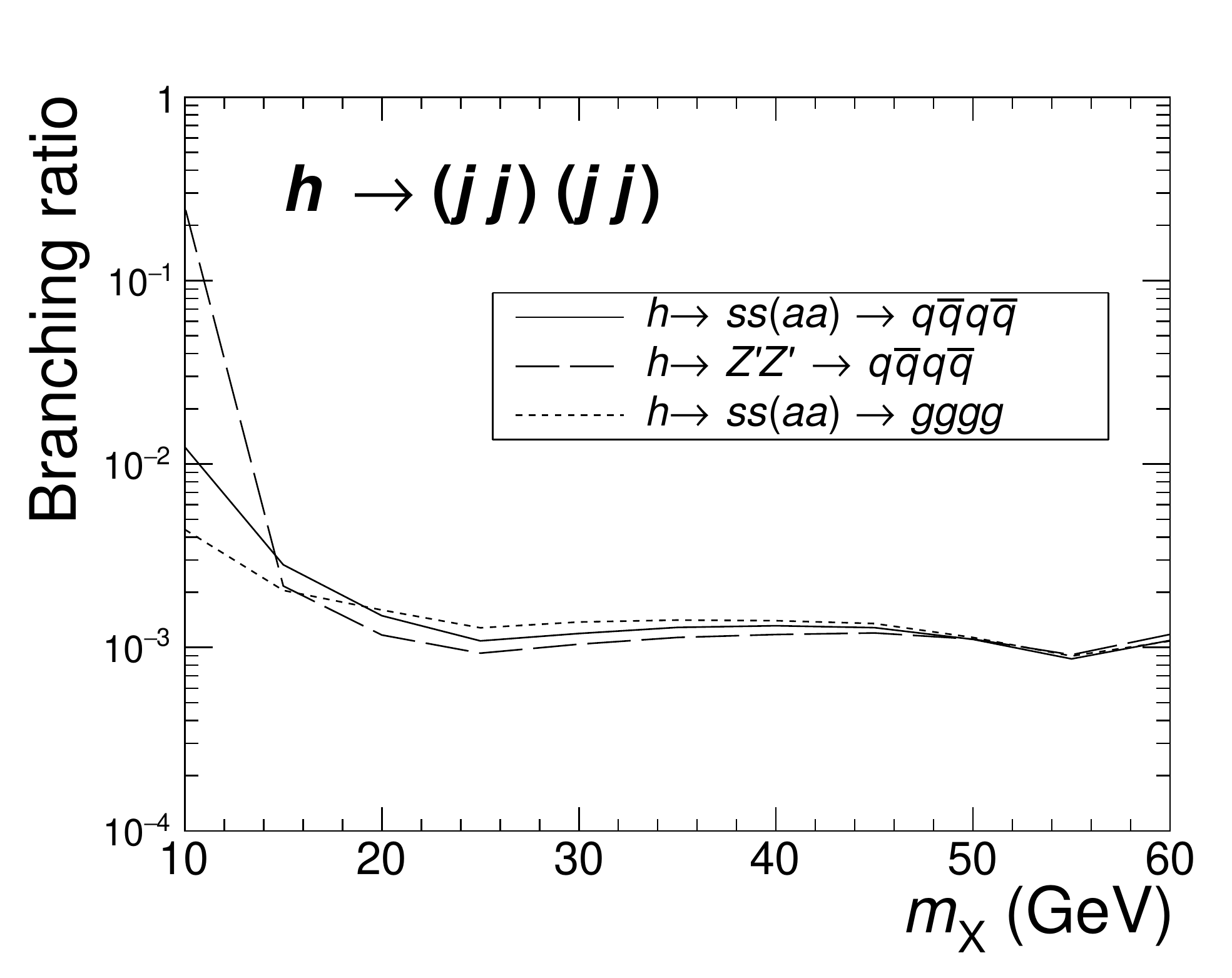}
\includegraphics[width=8cm]{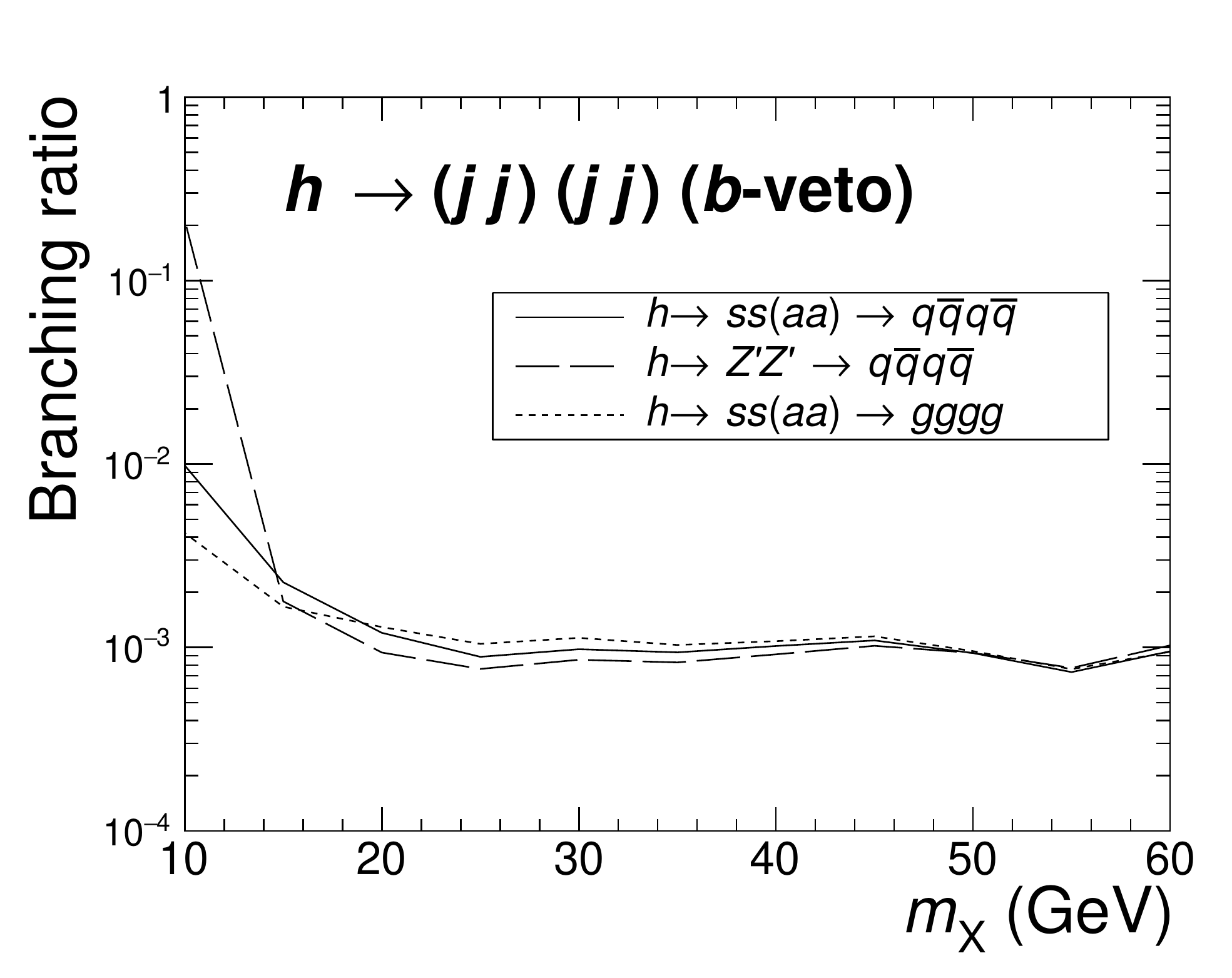}
\figcaption{\label{fig:brjjjj1}The 95\% C.L. exclusive bound of Br$(h\to (jj)(jj))$ 
(a) without and (b) with $b$-veto.}  
\end{center}

If the mediator decays to the light flavors ($u,d,s,c,g$) but not the bottom quark, a
$b$-veto could be used to suppress the SM background. In this case, we 
require that there is no $b$-tagged jet in the final state. The $b$-veto does not 
increase the sensitivity significantly, as shown in Fig.~\ref{fig:brjjjj1}(b), 
because the SM background is dominated by the 
$Z(h\to VV^*\to jjjj)$, which has a similar composition of quark flavors.

The future lepton collider with $5~\abi$ integrated luminosity could exclude $10^{-3}$ 
branching fractions of $h\to (jj)(jj)$ in a wide range of the mediator mass. We do not 
consider the mediator mass below 10~\gev\, as a different analysis strategy should then be 
taken because the jets are more likely to fail the separation cuts, manifested in the 
left-hand corner of Fig.~\ref{fig:zjjjj}.

\begin{center}
\includegraphics[width=8cm]{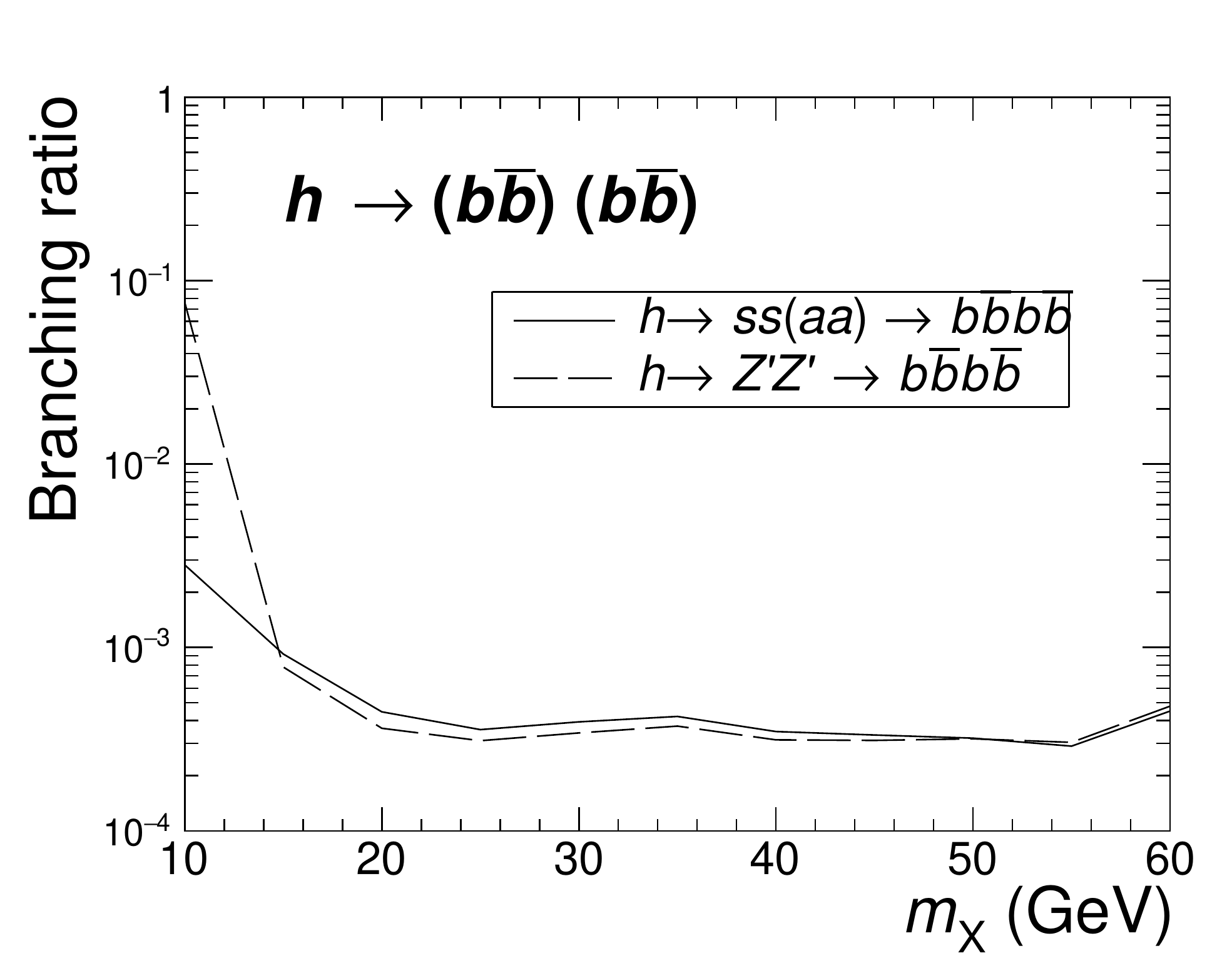}
\includegraphics[width=8cm]{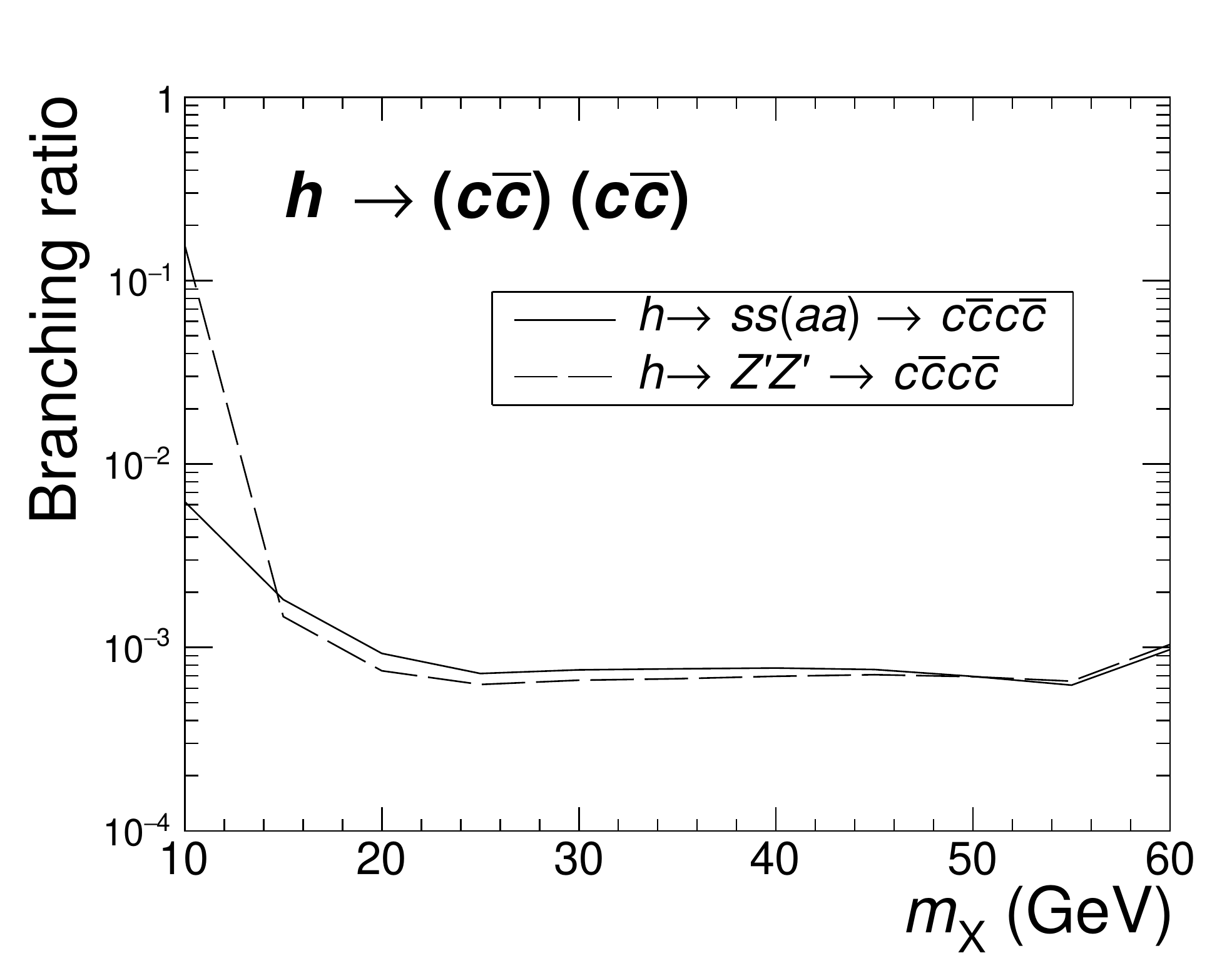}
\figcaption{\label{fig:brjjjj2}The 95\% C.L. exclusive bound on (a) Br$(h\to (b\bar b)(b\bar b))$ 
 and (b) Br$(h\to (c\bar c)(c\bar c))$.}  
\end{center}

The sensitivity in this channel is worse than in $h\to (jj)+\met$ and $h\to jj+\met$.
This is because we miss the information of the correct combination of jets in the 
final state. The wrong combination of jets in the final state from the SM background 
will mimic the signal.

For the case of di-bottom pair resonances $h\to (b\bar b)(b\bar b)$ and  di-charm pair 
resonances $h\to (c\bar c)(c\bar c)$, the simulation is nearly identical to the $h\to(jj)(jj)$ 
case.  We require at least three $b$-tagged jets and $c$-tagged jets in the final state, 
respectively. For the charm-tagging, we assume a tagging efficiency of 60\% and mis-tag 
rate from $b$-jets 15\% and light jets 10\%. The relatively larger fake rate assumed here 
leads to slightly worse sensitivity for $h\to (c\bar c)(c\bar c)$ when comparing with $h\to 
(b\bar b)(b\bar b)$. The results are shown in Fig.~\ref{fig:brjjjj2}. The future lepton collider 
with $5~\abi$ integrated luminosity could exclude  branching fractions of $h\to (b\bar b)(b\bar b)$ and 
$h\to (c\bar c)(c\bar c)$ down to $3\times10^{-4}\sim 4\times 10^{-4}$ and 
$7\times 10^{-4}\sim 9\times 10^{-4}$, respectively, in a wide range of the mediator mass.

\subsection{$h\to(\gamma\gamma)(\gamma\gamma)$}
For this scenario, we consider the (pseudo)scalar mediator.
In addition to the preliminary cuts, we require
there are at least four hard photons that satisfy 
\beq
E_\gamma>10~{\text{GeV}}.
\label{eq:aaaacut}
\eeq

The most important background from the SM is 
\beq
e^+e^-\to Z+4\gamma \nonumber
\eeq
with
\beq
Z\to \ell^+\ell^-.
\eeq

\begin{center}
\includegraphics[width=8cm]{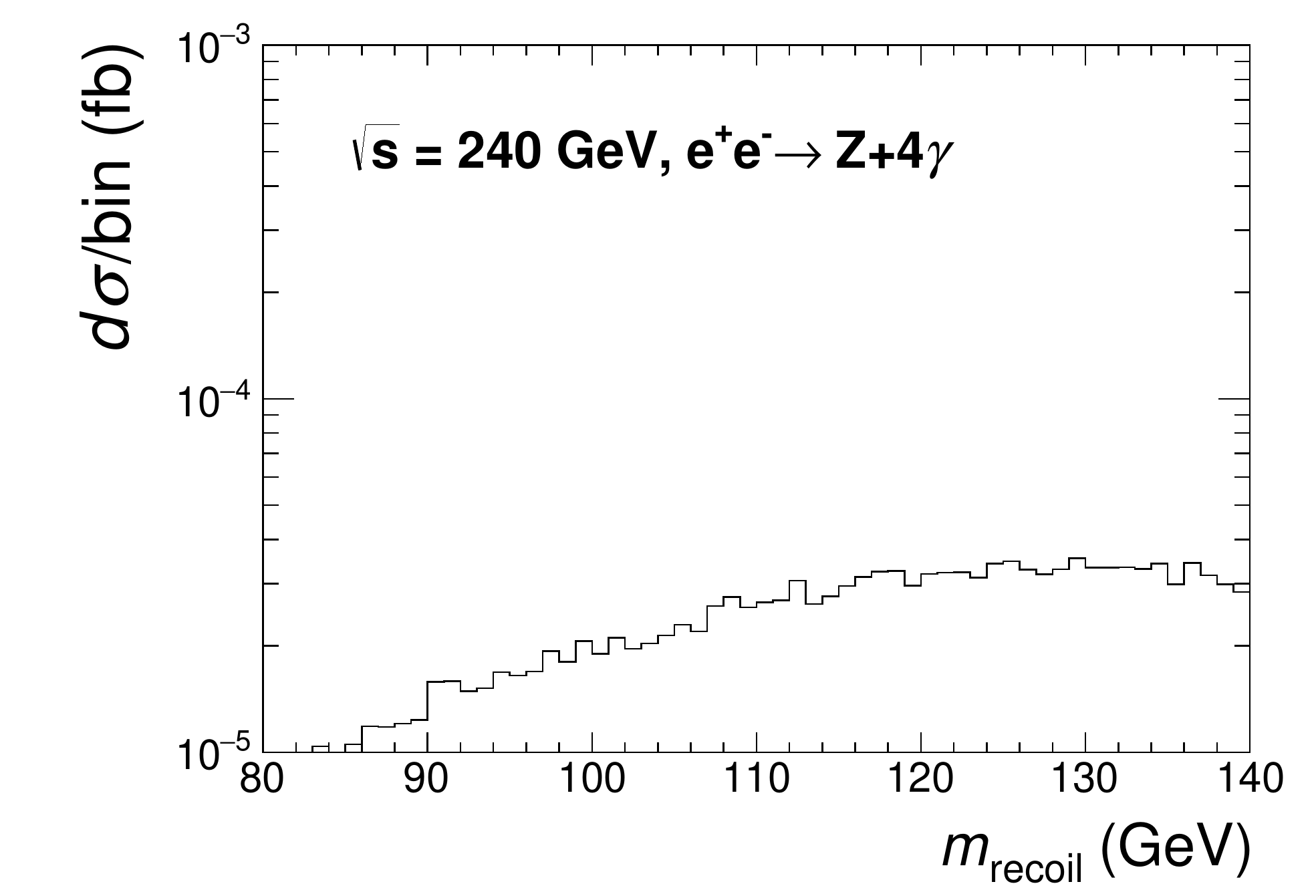}
\figcaption{\label{fig:zaaaa}The recoil mass distribution of the SM backgrounds for $Z+4\gamma$. 
 The $R_m$ cut and all of the pre-selection
cuts except for the recoil mass cut are applied. }  
\end{center}
We show the SM background distribution in the recoil mass after imposing all the 
other kinematical cuts in Fig.~\ref{fig:zaaaa}. The contribution from $e^+e^-\to Z+h+
2\gamma\to \ell^+\ell^-+4\gamma$ is highly suppressed by the $h\to\gamma\gamma$ 
decay branching ratio.~\footnote{The cross section of $e^+e^-\to Z+h\to \ell^+\ell^-+
2\gamma$ is less than 40~ab and is severely suppressed by two more hard photons 
in the final state, so it is neglected here.}
The SM $Z$ boson decaying into a four photon final state has a tiny 
branching ratio and thus  it does not contribute. Therefore,  there is no sizable resonance structure 
in the recoil mass distribution in the SM background of this channel, as shown in  this figure. 

Beyond the pre-selection cuts, the four photons should form two pairs of diphoton 
resonances with similar masses and thus we further require
\beq
R_m\equiv\min_{\sigma\in A_4}\left|\frac{m_{\gamma_{\sigma(1)}\gamma_{\sigma(2)}}
-m_{\gamma_{\sigma(3)}\gamma_{\sigma(4)}}}{m_{\gamma_{\sigma(1)}\gamma_{\sigma(2)}}
+m_{\gamma_{\sigma(3)}\gamma_{\sigma(4)}}}\right|<0.1
\eeq
for a signal event. The SM background is $0.14\times 10^{-3}$ fb.
Around three signal events are needed to be excluded at 95\% C.L. for $5~\abi$ integrated luminosity.

\begin{center}
\tabcaption{\label{tab:aaaa}The cut acceptances for the $h\to (\gamma\gamma)(\gamma\gamma)$ and 
the derived 95\% C.L. limits on decay branching ratio 
Br$(h\to (\gamma\gamma)(\gamma\gamma))$ with $5~\abi$ integrated luminosity for 
various masses of the scalar mediator.}
\footnotesize
\begin{tabular*}{80mm}{c@{\extracolsep{\fill}}cccccc}
\toprule $m_{\text{med}}$ (GeV)&10& 20 & 25 &30&50\\
\hline
$sF_{\mu\nu}F^{\mu\nu}, aF_{\mu\nu}\tilde F^{\mu\nu}$
&7.8\%&33\%&37\%&41\%&60\%\\
Br$(h\to (\gamma\gamma)(\gamma\gamma))$ ($10^{-4}$)
&5.1&1.2&1.1&0.97&0.66\\
\bottomrule
\end{tabular*}
\vspace{0mm}
\end{center}

We tabulate the 
signal efficiency for various intermediate masses and show the derived 95\% C.L. limits on 
this Higgs exotic branching fraction Br$(h\to (\gamma\gamma)(\gamma\gamma))$ in Table 
\ref{tab:aaaa}. The signal acceptance for this analysis increases as the mediator 
mass increases, as a result of the requirement on the photon energy in Eq.~(\ref{eq:aaaacut}). 
The limits can reach up to $4.7\times 10^{-4}$ on the exotic branching fractions of this channel 
for a mediator mass of 50~GeV. 

\subsection{$h\to(jj)(\gamma\gamma)$}
For this scenario, we consider the (pseudo)scalar mediator.
In additional to the preliminary cuts, we require
there are at least two hard photons and two hard jets that satisfy 
\beq
E_{\gamma,j}>10~{\text{GeV}}.
\label{eq:jjaacut}
\eeq

The most important background from the SM is 
\beq
e^+e^-\to Z+2\gamma+2j \nonumber
\eeq
with
\beq
Z\to \ell^+\ell^-.
\eeq
The contribution from $e^+e^-\to Z+h\to \ell^+\ell^-+b\bar b\to\ell^+\ell^-+b\bar b+2\gamma$ 
is suppressed by the fine structure constant and the separation cut on $y$ in Eq.~(\ref{eq:separationcut}).
We show the SM background distribution in the recoil mass with the all but the recoil mass cut in Fig.~\ref{fig:zjjaa}.
The background is mostly from $e^+e^-\to ZZ+2\gamma$ where the photons are from 
the ISR. There is also a sizable contribution from $e^+e^-\to ZZ$ where one of the $Z$ boson 
decays into a $q\bar q$ final state and the photons are from the final state radiation from the jets. We 
can find a small peak at the $Z$-pole in the recoil mass distribution in this figure.
\begin{center}
\includegraphics[width=8cm]{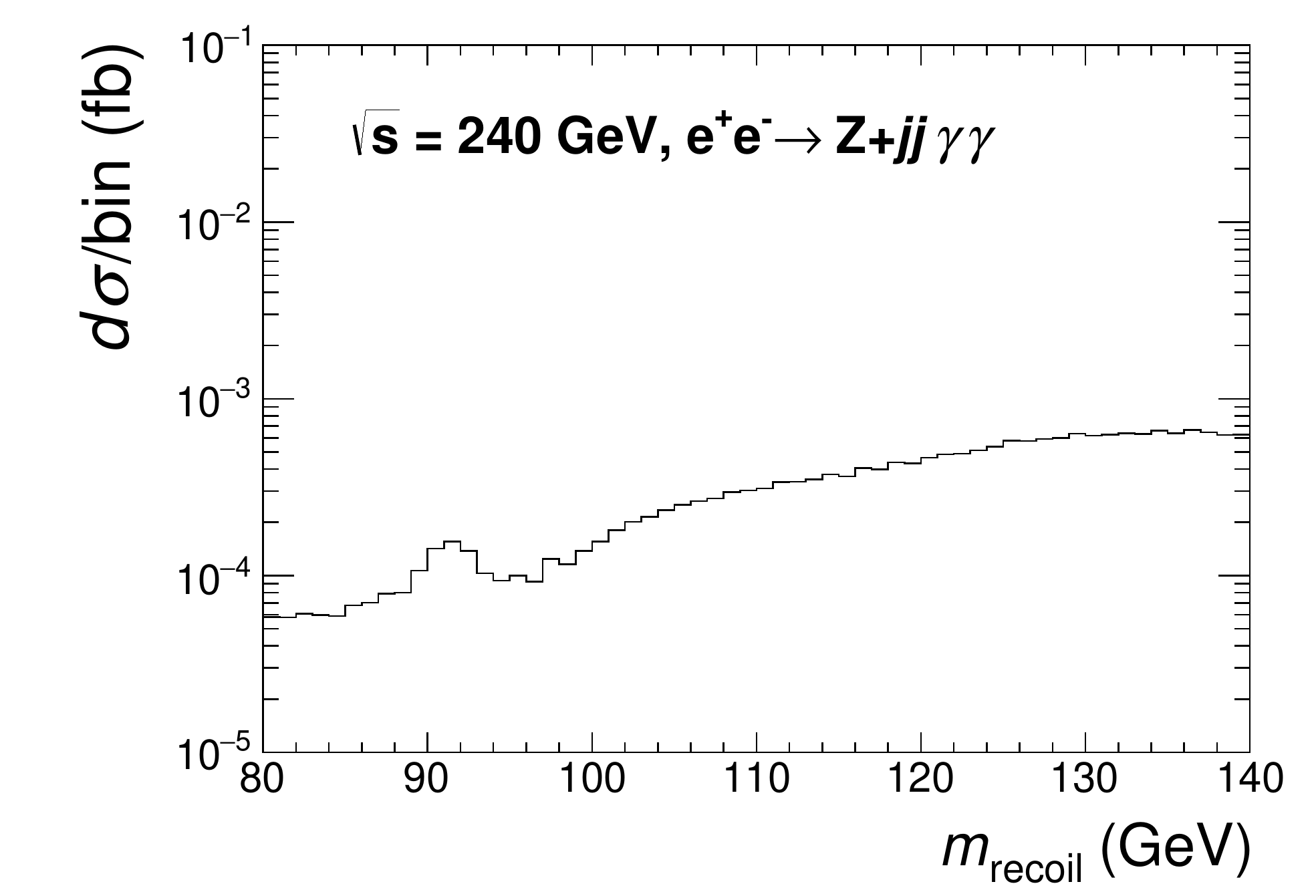}
\figcaption{\label{fig:zjjaa}The recoil mass distribution of the SM backgrounds for $Z+2\gamma+2j$.
All but the recoil mass cut are applied.}  
\end{center}

We also require the dijet invariant mass and diphoton invariant mass to be close in value,
\beq
R_m\equiv\frac{\left|m_{jj}
-m_{\gamma\gamma}\right|}{m_{jj}
+m_{\gamma\gamma}}<0.1
\eeq
to further suppress the background.

The SM background is $0.31\times 10^{-3}$ fb. 
Around four signal events are needed to be excluded at 95\% C.L. for $5~\abi$ integrated luminosity.

\begin{center}
\tabcaption{\label{tab:jjaa}The cut acceptances for the $h\to (jj)(\gamma\gamma)$ and the derived 
95\% C.L. limits on decay branching ratio 
Br$(h\to (jj)(\gamma\gamma))$ with $5~\abi$ integrated luminosity for various masses of the scalar mediator.}
\footnotesize
\begin{tabular*}{80mm}{c@{\extracolsep{\fill}}cccccc}
\toprule $m_{\text{med}}$ (GeV)&10& 20 & 25 &30&50\\
\hline
cut acceptance  
&10\%&34\%&39\%&44\%&66\%\\
Br$(h\to (jj)(\gamma\gamma))$ $(10^{-4})$
&5.3&1.6&1.4&1.2&0.81\\
\bottomrule
\end{tabular*}
\vspace{0mm}
\end{center}

We tabulate the 
signal efficiency for various intermediate masses and show the derived 95\% C.L. 
limits on this Higgs exotic branching fraction Br$(h\to (jj)(\gamma\gamma))$ in Table 
\ref{tab:jjaa}. We can see the signal selection efficiency for this analysis increases as 
the mediator mass increases, as a result of the requirement on  final state particle energies 
in Eq.~(\ref{eq:jjaacut}). The limits can reach up to $5.6\times 10^{-4}$ on the exotic 
branching fractions of this channel for a mediator mass of 50~GeV. 

\section{Summary and outlook}
We summarize the set of Higgs exotic decays in Table~\ref{tab:summary}, including current 
collections and projections of LHC constraints, and limits from our study for the CEPC, ILC and 
FCC-$ee$. For the LHC constraints, we tabulate both the current limits and projected limits 
on these exotic decay channels from various references. Due to the subtleties in the projections 
at the LHC from possible systematics, we only quote the projected limits from studies with 
highest integrated luminosity and label them accordingly, instead of naively scaling these 
results to the projected luminosity of  (HL-)LHC. The current LHC limits are collected in the 
second column, and the square brackets represent 7 and 8 TeV LHC results alone. The projections 
for HL-LHC are collected in the third column, where the limits for 100~$\fbi$~ and 300~$\fbi$ are 
shown in parentheses and square brackets respectively. The reference for the current limits and projected 
limits are also included. For projections on the reaches of HL-LHC, it should be understood that 
they are from  Ref.~\cite{Curtin:2013fra} unless specified otherwise. The limits for the case of the 
ILC and FCC-$ee$ are extrapolated from our CEPC analyses in Section~3 following 
the running scenario discussion in Section~2, assuming the reach is dominated by 
statistics. The decay channels involving $\tau$-leptons are extrapolated and validated by our 
preliminary simulation, assuming a 40\% $\tau$-tagging efficiency with more background considered. 
We mark these extrapolated limits with an asterisk.

\end{multicols}
\begin{center}
\tabcaption{\label{tab:summary}The current and projected limits on selected Higgs exotic 
decay modes for the (HL-)LHC, CEPC, ILC, and FCC-$ee$. Throughout our analysis, we assume 
SM production rates for the Higgs boson. The current LHC limits are collected in the second 
column, with the square brackets indicating 7 and 8 TeV LHC results alone. The projections 
for the HL-LHC are collected in the third column, where the limits for 100 $\fbi$   and 300 $\fbi$ 
alone are shown in parentheses and square brackets respectively. The references for the current limits 
and projected limits are included; for projections on the HL-LHC limits we omit the references 
if they are from Ref.~\cite{Curtin:2013fra}. We also omit the references for projections on the 
future lepton collider programs if the results are from this study. Limits extracted from related 
searches with reasonable assumptions are marked with asterisks following the results with details 
explained in the text. For more details about the benchmark parameter choices for some of the 
decay modes, see discussion in the main text. }
\footnotesize
\begin{tabular*}{117.5mm}{c@{\extracolsep{\fill}}cccccc}
\toprule Decay & \multicolumn{5}{c}{95\% C.L. limit on Br}\\ \cline{2-6} 
Mode & LHC & HL-LHC & CEPC & ILC & FCC-$ee$\\
\hline\hline
$\met$& 0.23~\cite{Aad:2015pla,Khachatryan:2016whc} & 0.056~
\cite{CMS-NOTE-2013-002,ATL-PHYS-PUB-2013-014,CMS-DP-2016-064} 
& 0.0028~\cite{CEPCpreCDR} & 0.0025~\cite{Fujii:2015jha} & 0.005~\cite{Gomez-Ceballos:2013zzn}\\
\hline
$ (b\bar b)+\met$ & -- & [0.2] & \scidgts{1}{-4} & \scidgts{2}{-4} & \scidgts{5}{-5}\\
$ (jj)+\met$& -- & -- & \scidgts{5}{-4} & \scidgts{5}{-4} & \scidgts{2}{-4}\\
$ (\tau^+\tau^-)+\met$ & -- &  [1] & $\scidgts{8}{-4}$* & \scidgts{1}{-3} & \scidgts{3}{-4}\\
\hline
$ b\bar b+\met$& -- & [0.2]~\cite{Huang:2013ima} & \scidgts{3}{-4} & \scidgts{4}{-4} & \scidgts{1}{-4}\\
$ jj+\met$& -- & -- & \scidgts{5}{-4} & \scidgts{7}{-4} & \scidgts{2}{-4} \\
$ \tau^+\tau^-+\met$& -- & -- & \scidgts{8}{-4}* & \scidgts{1}{-3} & \scidgts{3}{-4}\\
\hline
$ (b\bar b)(b\bar b)$& 1.7~\cite{Aaboud:2016oyb} & (0.2) 
& \scidgts{4}{-4} & \scidgts{9}{-4} & \scidgts{3}{-4}\\
$(c\bar c)(c\bar c)$& -- & (0.2) 
& \scidgts{8}{-4}&\scidgts{1}{-3} & \scidgts{3}{-4}\\
$(jj)(jj)$& -- & $[0.1]$&
 \scidgts{1}{-3} & \scidgts{2}{-3} & \scidgts{7}{-4}\\
$ (b\bar b)(\tau^+\tau^-)$& [0.1]*~\cite{CMS-PAS-HIG-14-041}& [0.15] & \scidgts{4}{-4}* 
& \scidgts{6}{-4} & \scidgts{2}{-4}\\
$ (\tau^+\tau^-)(\tau^+\tau^-)$& [1.2]*~\cite{Aad:2015oqa} & $[0.2\sim0.4]$ 
& \scidgts{1}{-4}* & \scidgts{2}{-4} & \scidgts{5}{-5}\\
$(jj)(\gamma\gamma)$& -- & $[0.01]$ & \scidgts{1}{-4} & \scidgts{2}{-4} & \scidgts{3}{-5}\\
$(\gamma\gamma)(\gamma\gamma)$& [\scidgts{7}{-3}]~\cite{Aad:2015bua}& $\scidgts{4}{-4}*$&
\scidgts{1}{-4} & \scidgts{1}{-4} & \scidgts{3}{-5}\\
\bottomrule
\end{tabular*}
\vspace{0mm}
\end{center}
\begin{multicols}{2}

Similar to the lepton collider, the LHC reach can depend on model parameters. 
While a comparison of the reach throughout the full parameter space is possible, 
it is tedious and not illuminating. We choose to focus on comparison for particular 
benchmark points, which is good enough to demonstrate the qualitative difference 
between the LHC and future lepton colliders. These choices are to make the limits 
more or less representative of the intermediate~(``average'') limits across wide parameter 
regions without being in the extremely good or bad kinematical points.

The LHC reach for $(b\bar b)(\tau^+\tau^-)$ and $(\tau^+\tau^-)(\tau^+\tau^-)$ is
extrapolated from the $(b\bar b)(\mu^+\mu^-)$ and $(\tau^+\tau^-)(\mu^+\mu^-)$ 
searches, respectively, assuming $Br(a\to \tau^+\tau^-)/Br(a\to \mu^+\mu^-)=
m_\tau^2/m_\mu^2$. For the process of $h\to a a \to (b\bar b) (\mu^+\mu^-)$ with 
$m_a=30~\gev$, the HL-LHC projected sensitivity is $5\times 10^{-5}$~\cite{Curtin:2014pda}, 
translating into a limit on $h\to a a \to (b\bar b) (\tau^+\tau^-)$ of $0.02$, better than the 
projected direct search limit on this channel from Ref.~\cite{Curtin:2013fra}.

Photons are one of the ``clean'' objects in the LHC collider environment and the large 
statistics at the HL-LHC can constrain decays with photons to good precision. However, 
these is some discrepancy between the theoretical projections in Ref.~\cite{Curtin:2013fra} 
of $3\times 10^{-5}$ at $300~\fbi$, and the extrapolation from the current limit at the 8 TeV 
LHC~\cite{Aad:2015bua}. We choose to use the more conservative extrapolation by us and 
thus put an asterisk after the projected limit in this summary table.

 For the decay topologies involving intermediate resonant particles, we choose the 
 intermediate particle to be a pseudoscalar with a mass of $30~\gev$ as a benchmark, 
 which applies to the $(b\bar b)(b\bar b)$, $(c\bar c)(c\bar c)$, $(jj)(jj)$, $(b\bar b)(\tau^+\tau^-)$, 
 $(\tau^+\tau^-)(\tau^+\tau^-)$, $(jj)(\gamma\gamma)$, and $(\gamma\gamma)(\gamma\gamma)$ 
 decay channels. For a decay topology of $h\to 2 \to 3 \to 4$ where intermediate resonances 
 are involved, we choose the lightest stable particle mass to be $10~\gev$, the mass splitting 
 to be $40~\gev$ and the intermediate resonance mass to be 10~\gev, which applies to 
 $(b\bar b)+\met$, $(jj)+\met$, $(\tau^+\tau^-)+\met$. For a decay topology of $h\to 2 \to (1+3)$, 
 we choose the lightest stable particle mass to be $10~\gev$ and the mass splitting to be 
 $40~\gev$, which applies to $b\bar b+\met$, $jj+\met$, $\tau^+\tau^-+\met$.  For the Higgs 
 invisible decays, we take the best limits in the running scenario ECFA16-S2 amongst the $Zh$ 
 associated production and VBF search channels~\cite{CMS-NOTE-2013-002,
 ATL-PHYS-PUB-2013-014,CMS-DP-2016-064}. 
 
 For the Higgs invisible decays at lepton colliders, we quote the limits from current studies
 \cite{CEPCpreCDR,Fujii:2015jha,Gomez-Ceballos:2013zzn}. These limits do not depend 
 on the invisible particle mass using the recoil mass technique at lepton colliders.

\end{multicols}
\begin{center}
\includegraphics[width=16cm]{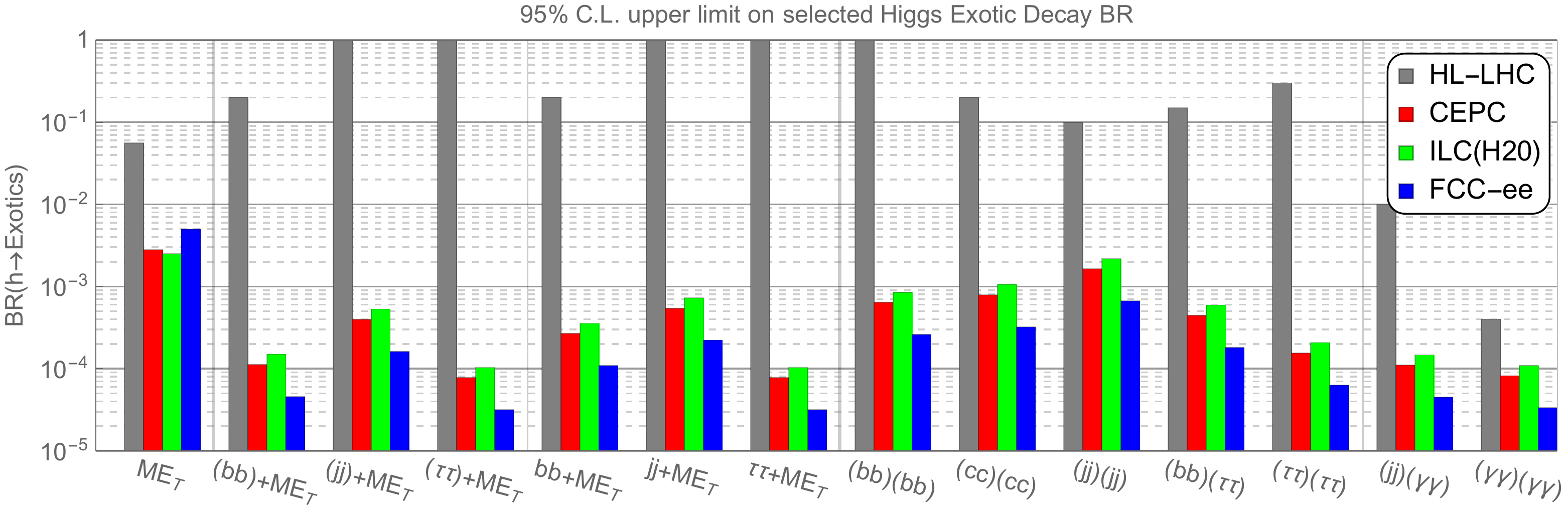}
\figcaption{\label{fig:summary} The 95\% C.L. upper limit on selected Higgs exotic decay 
branching fractions at HL-LHC, CEPC, ILC and FCC-$ee$. The benchmark parameter choices 
are the same as in Table~\ref{tab:summary}. We put several vertical lines in this figure to 
divide different types of Higgs exotic decays.}  
\end{center}
\begin{multicols}{2}

From this summary in Table~\ref{tab:summary} and the corresponding Fig.~\ref{fig:summary}, 
we can clearly see the improvement in exotic decays from the lepton collider Higgs factories. 
These exotic Higgs decay channels are selected such that they are hard to be constrained at 
the LHC but important for probing BSM decays of the  Higgs boson. The improvements on the 
limits of the Higgs exotic decay branching fractions vary from one to four orders of magnitude 
for these channels. The lepton colliders can improve the limits on the Higgs invisible decays 
beyond the HL-LHC projection by one order of magnitude, reaching the SM invisible decay 
branching fraction of 0.12\% from $h\to ZZ^*\to \nu\bar \nu \nu\bar \nu$~\cite{Heinemeyer:2013tqa}. 
For the Higgs exotic decays into hadronic particle plus missing energy, $(b\bar b)+\met$, 
$(jj)+\met$ and $(\tau^+\tau^-)+\met$, the future lepton colliders improve on the HL-LHC sensitivity 
for these channels by roughly four orders of magnitude. This great advantage benefits a lot from 
low QCD background and the Higgs tagging from recoil mass technique at future lepton colliders. 
As for the Higgs exotic decays without missing energy, the improvement varies between two to 
three orders of magnitude, except for the one order of magnitude improvement for the 
$(\gamma\gamma)(\gamma\gamma)$ channel. Being able to reconstruct the Higgs mass from 
the final state particles at the LHC does provide additional signal-background discrimination 
power and hence the future lepton colliders improvement on Higgs exotic decays without missing 
energy is less impressive than for those with missing energy. Furthermore, as discussed earlier, leptons 
and photons are relatively clean objects at the LHC and the sensitivity at the LHC on these 
channels will be very good. Future lepton colliders complement the HL-LHC for hadronic 
channels and channels with missing energies.

There are many more investigations to be carried out under the theme of Higgs exotic decays. 
For our study, we take the cleanest channel of $\ee\to ZH$ with $Z\to \ell^+\ell^-$ and $h\to
${exotics} up to four-body final state, but further inclusion of the hadronic decaying spectator 
$Z$-boson and even invisible decays of the $Z$-boson would definitely improve the statistics 
and consequently result in better limits. As a first attempt to evaluate the Higgs exotic decay 
program at future lepton colliders, we do not include the case of very light intermediate particles 
whose decay products will be collimated, but postpone this for future study when the detector 
performance is more clearly defined. There are many more exotic Higgs decay modes to consider, 
such as Higgs decaying to a pair of intermediate particles with un-even masses~\cite{Curtin:2013fra}, 
Higgs CP property measurements from its decay differential distributions~\cite{Harnik:2013aja,
Li:2015kxc,Craig:2015wwr,Li:2016zzh},  flavor violating decays, decays to light quarks~\cite{Gao:2016jcm}, 
decays into meta-stable particles, and complementary Higgs exotic productions~\cite{Cao:2015iua}. 
Our work is a first systematic study evaluating the physics potential of future lepton colliders to probe 
Higgs exotic decays. More work will be needed to obtain a complete picture. 

\acknowledgments {LTW would like to thank Matt Strassler for useful discussions, and the Institute of High Energy Physics 
in Beijing for hospitality. ZL and LTW would like to thank the Kavli Institute for Theoretical Physics for hospitality. }

\end{multicols}

\vspace{-1mm}
\centerline{\rule{80mm}{0.1pt}}
\vspace{2mm}

\begin{multicols}{2}

\end{multicols}

\clearpage
\end{CJK*}

\begin{thebibliography}{90}

\vspace{3mm}

\bibitem{Aad:2012tfa}G.~Aad et~al (ATLAS Collaboration), Phys. Lett. B, {\bf716}: 1--29 (2012)

\bibitem{Chatrchyan:2012xdj}S.~Chatrchyan et~al (CMS Collaboration), Phys. Lett. B, {\bf716}: 30--61(2012)

\bibitem{Flores:1982pr}R.~A. Flores and M.~Sher, Annals Phys., {\bf 148}: 95 (1983)

\bibitem{Gunion:1984yn}J.~F. Gunion and H.~E. Haber, Nucl. Phys. B, {\bf272}: 1 (1986)

\bibitem{Djouadi:2005gj}A.~Djouadi, Phys. Rept., {\bf 459}: 1--241 (2008)

\bibitem{Gripaios:2009pe}B.~Gripaios, A.~Pomarol, F.~Riva et~al., JHEP, {\bf 04}: 70 (2009)

\bibitem{ArkaniHamed:1998nn}N.~Arkani-Hamed, S.~Dimopoulos and G.~R. Dvali, Phys. Rev. D, {\bf59}: 086004 (1999)

\bibitem{Randall:1999ee}L.~Randall and R.~Sundrum, Phys. Rev. Lett., {\bf 83}: 3370--3373 (1999)

\bibitem{Randall:1999vf}L.~Randall and R.~Sundrum, Phys. Rev. Lett., {\bf 83}: 4690--4693 (1999)

\bibitem{Georgi:1974sy}H.~Georgi and S.~L. Glashow, Phys. Rev. Lett., {\bf 32}: 438--441 (1974)

\bibitem{Dawson:2013bba}S.~Dawson et~al., arXiv:1310.8361

\bibitem{CMS-NOTE-2013-002}CMS Collaboration, arXiv:1307.7135

\bibitem{CMS-DP-2016-064}https://cds.cern.ch/record/2221747/files/DP2016\_064.pdf, retrieved 4th October 2016

\bibitem{ATL-PHYS-PUB-2013-014}http://cds.cern.ch/record/1611186/files/ATL-PHYS-PUB-2013-014.pdf, retrieved 17th October 2013

\bibitem{ATL-PHYS-PUB-2014-016}http://cds.cern.ch/record/1956710/files/ATL-PHYS-PUB-2014-016.pdf, retrieved 21st October 2014

\bibitem{CEPCpreCDR}http://cepc.ihep.ac.cn/preCDR/main\_preCDR.pdf, retrieved 4th May 2015

\bibitem{Fujii:2015jha}K.~Fujii et~al., arXiv:1506.05992

\bibitem{Gomez-Ceballos:2013zzn}M.~Bicer et~al (TLEP Design Study Working Group Collaboration), JHEP, {\bf 01}: 164 (2014)

\bibitem{Weinberg:1978kz}S.~Weinberg, Physica A, {\bf96}: 327--340 (1979)

\bibitem{Buchmuller:1985jz}W.~Buchmuller and D.~Wyler, Nucl. Phys. B, {\bf268}: 621--653 (1986)

\bibitem{Grzadkowski:2010es}B.~Grzadkowski, M.~Iskrzynski, M.~Misiak et~al., JHEP, {\bf 10}: 85 (2010)

\bibitem{Ellis:2014jta}J.~Ellis, V.~Sanz and T.~You, JHEP, {\bf 03}: 157 (2015)

\bibitem{Biekoetter:2014jwa}A.~Biek\"{o}tter, A.~Knochel, M.~Kr\"{a}mer et~al., Phys. Rev. D, {\bf91}: 055029 (2015)

\bibitem{Contino:2016jqw}R.~Contino, A.~Falkowski, F.~Goertz et~al., JHEP, {\bf 07}: 144 (2016)

\bibitem{Curtin:2013fra}D.~Curtin, R.~Essig, S.~Gori et~al., Phys. Rev. D, {\bf90}: 075004 (2014)

\bibitem{deFlorian:2016spz}D.~de~Florian et~al., arXiv:1610.07922

\bibitem{Patt:2006fw}B.~Patt and F.~Wilczek, arXiv: hepph/0605188

\bibitem{Arbey:2012na}A.~Arbey, M.~Battaglia and F.~Mahmoudi, Eur. Phys. J. C, {\bf 72}: 2169 (2012)
  
\bibitem{BhupalDev:2012ru}P.~S. Bhupal~Dev, S.~Mondal, B.~Mukhopadhyaya et~al., JHEP, {\bf 09}: 110 (2012)

\bibitem{Han:2013gba}T.~Han, Z.~Liu and A.~Natarajan, JHEP, {\bf 11}: 8 (2013)

\bibitem{Banerjee:2013fga}S.~Banerjee, P.~S.~B. Dev, S.~Monda et~al., JHEP, {\bf 10}: 221 (2013)

\bibitem{Buckley:2013sca}M.~R. Buckley, D.~Hooper and J.~Kumar, Phys. Rev. D, {\bf88}: 063532 (2013)

\bibitem{Hagiwara:2013qya}K.~Hagiwara, S.~Mukhopadhyay and J.~Nakamura, Phys. Rev. D, {\bf89}: 015023 (2014)

\bibitem{Belanger:2013pna}G.~B\'{e}langer, G.~Drieu La~Rochelle, B.~Dumont et~al., Phys. Lett. B, {\bf726}: 773--780 (2013)

\bibitem{Pierce:2013rda}A.~Pierce, N.~R. Shah and K.~Freese, arXiv:1309.7351

\bibitem{Cao:2013mqa}J.~Cao, C.~Han, L.~Wu et~al., JHEP, {\bf 05}: 56 (2014)

\bibitem{Han:2014nba}T.~Han, Z.~Liu and S.~Su, JHEP, {\bf 08}: 93 (2014)

\bibitem{Huang:2014cla}J.~Huang, T.~Liu, L.-T. Wang et~al., Phys. Rev. D, {\bf90}: 115006 (2014)

\bibitem{Huang:2013ima}J.~Huang, T.~Liu, L.-T. Wang et~al., Phys. Rev. Lett., {\bf112}: 221803 (2014)

\bibitem{Fan:2011yu}J.~Fan, M.~Reece and J.~T. Ruderman, JHEP, {\bf 11}: 12 (2011)

\bibitem{Xiu:2015tha}Q.~Xiu, H.~Zhu, X.~Lou et~al., Chin. Phys. C, {\bf40}: 053001 (2016)

\bibitem{Greco:2016izi}M.~Greco, T.~Han and Z.~Liu, Phys. Lett. B, {\bf763}: 409--415 (2016)

\bibitem{Essig:2013lka}R.~Essig et~al., arXiv:1311.0029

\bibitem{Fox:2011qd}P.~J. Fox, J.~Liu, D.~Tucker-Smith et~al., Phys. Rev. D, {\bf84}: 115006 (2011)

\bibitem{Chacko:2005pe}Z.~Chacko, H.-S. Goh and R.~Harnik, Phys. Rev. Lett., {\bf 96}: 231802 (2006)

\bibitem{Sun:2016bel}Q.-F. Sun, F.~Feng, Y.~Jia et~al., arXiv:1609.03995

\bibitem{Gong:2016jys}Y.~Gong, Z.~Li, X.~Xu et~al., arXiv:1609.03955

\bibitem{Alwall:2014hca}J.~Alwall, R.~Frederix, S.~Frixione et~al., JHEP, {\bf1407}: 79 (2014)

\bibitem{Aad:2015pla}G.~Aad et~al (ATLAS Collaboration), JHEP, {\bf11}: 206 (2015)

\bibitem{Khachatryan:2016whc}V.~Khachatryan et~al (CMS Collaboration), arXiv:1610.09218

\bibitem{Aaboud:2016oyb}M.~Aaboud et~al (ATLAS Collaboration), Eur. Phys. J. C, {\bf76}: 605 (2016)

\bibitem{CMS-PAS-HIG-14-041}https://cds.cern.ch/record/2135985/files/HIG-14-041-pas.pdf, retrieved 3rd March 2016

\bibitem{Aad:2015oqa}G.~Aad et~al (ATLAS Collaboration), Phys. Rev. D, {\bf92}: 052002 (2015)

\bibitem{Aad:2015bua}G.~Aad et~al (ATLAS Collaboration), Eur. Phys. J. C, {\bf76}: 210 (2016)

\bibitem{Curtin:2014pda}D.~Curtin, R.~Essig and Y.-M. Zhong, JHEP, {\bf06}: 25 (2015)

\bibitem{Heinemeyer:2013tqa}J.~R. Andersen et~al., arXiv:1307.1347

\bibitem{Harnik:2013aja}R.~Harnik, A.~Martin, T.~Okui et~al., Phys. Rev. D, {\bf88}: 076009 (2013)

\bibitem{Li:2015kxc}G.~Li, H.-R. Wang and S.-h. Zhu, Phys. Rev. D, {\bf93}: 055038 (2016)

\bibitem{Craig:2015wwr}N.~Craig, J.~Gu, Z.~Liu et~al., JHEP, {\bf03}: 50 (2016)

\bibitem{Li:2016zzh}G.~Li, Y.-n. Mao, C.~Zhang et~al, arXiv:1611.08518

\bibitem{Gao:2016jcm}J.~Gao, arXiv:1608.01746

\bibitem{Cao:2015iua}Q.-H. Cao, H.-R. Wang and Y.~Zhang, Chin. Phys. C, {\bf39}: 113102 (2015)

\bibitem{CMS-PAS-HIG-14-024}https://cds.cern.ch/record/1983181/files/HIG-14-024-pas.pdf, retrieved 27th January 2015

\end{thebibliography}
\end{document}